\documentstyle[epsfig]{bpl}

\begin{document}
\setcounter{page}{1}

~\\
%\bpl  2009\\
%BPL, {\bf 8} (2), pp. 105 - 109  (2000) \vskip14mm

\title{The Physics Prospects for CLIC}

\author{John ELLIS \\Theory Division, Physics Department, CERN,\\
CH-1211 Geneva 23, Switzerland\\
%[5mm]
%M. Ar{\i}k \\
%Physics Department, Bo§azi‡i University,\\
%Bebek, ˜stanbul, TURKEY.
%\\and\\
%Feza Grsey Institute,\\ P. O. Box 6, 81220\,-\,€engelk"y, ˜stanbul, TURKEY.
}

\maketitle

%\centerline{(received 2 January 2000 ; accepted 21 March 2000)}
\centerline{Talk presented at the International Conference on Particle Physics}
\centerline{held {\it in memoriam} for Engin Ar{\i}k and her colleagues}
\centerline{Bo{\g}azi{\c c}i University, Istanbul, Turkey, Oct. 27 to 31, 2008}
\centerline{~}
\centerline{CERN-PH-TH/2008-216}

\abstract{Following a brief outline of the CLIC project, this talk summarizes some of
the principal motivations for an $e^+ e^-$ collider with $E_{CM} = 3$~TeV. It is shown by 
several examples that CLIC would represent a significant step beyond the LHC and ILC in
its capabilities for precision measurements at high energies. It would make possible
a complete study of a light Higgs boson, including rare decay modes, and would
provide a unique tool to study a heavy Higgs boson.
CLIC could also complete the studies of supersymmetric spectra, if sparticles are
relatively light, and discover any heavier sparticles. It would also enable
deeper probes of extra dimensions, new gauge bosons and excited quarks or leptons.
CLIC has unique value to add to experimental particle physics, whatever the LHC
discovers.}\ea

%  Use the following example to insert EPS figures
%
%\begin{figure}[htbp]
%\centering
%\epsfig{file=asym.eps,width=5in,angle=0}
%\caption{This is a test figure.}
%\label{test_fig}
%\end{figure}
%

\section*{1 - The CLIC Project}

\begin{figure}[htbp]
\centering
{\epsfig{file=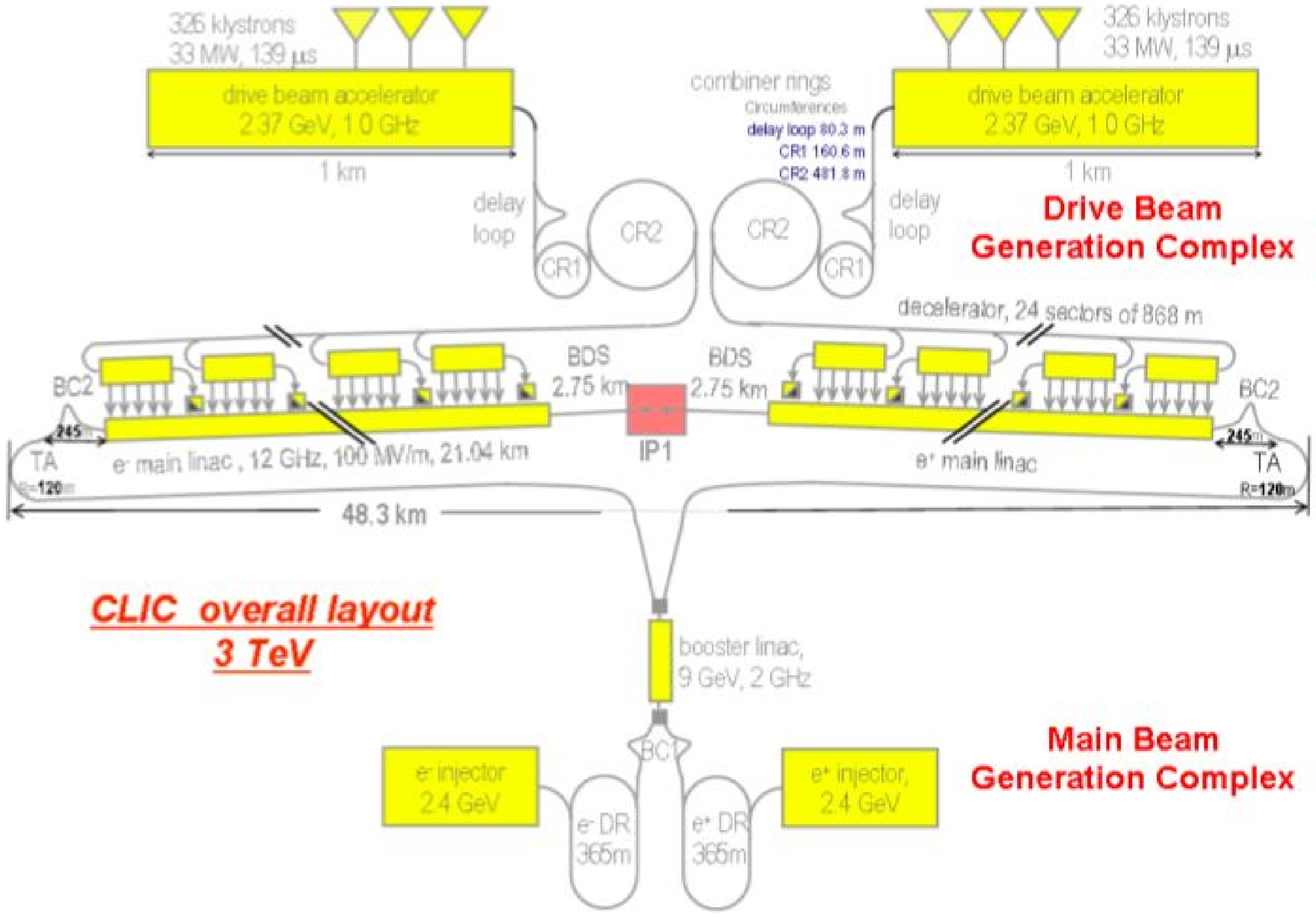,width=1.9in,angle=90}
\epsfig{file=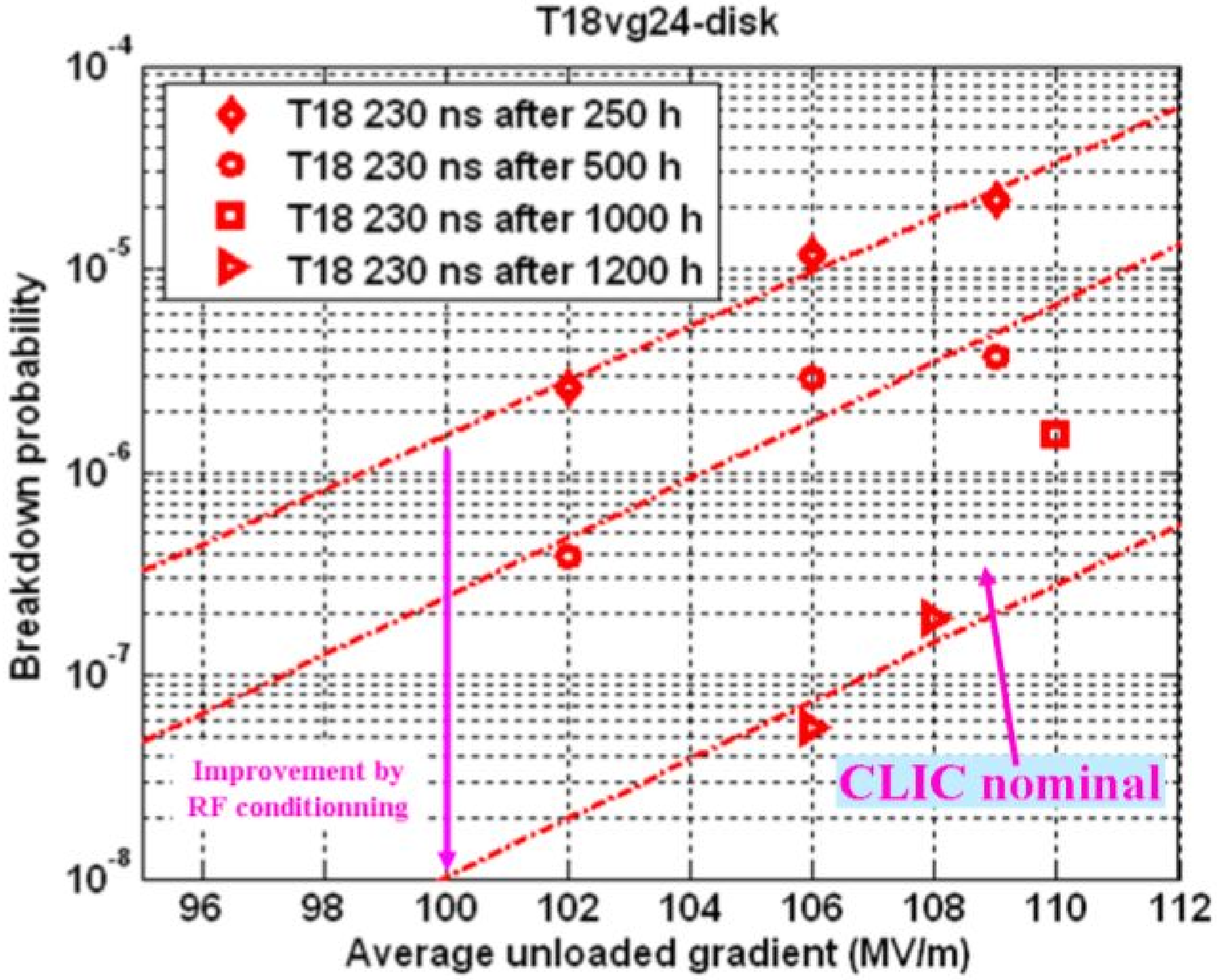,width=1.8in,angle=90}}
\caption{Left: The conceptual layout of CLIC at 3~TeV~\protect\cite{CLIC}. 
Right: the latest progress in
achieving high accelerating gradients in unloaded 12~GHz CLIC structure
T18~\protect\cite{accgrad}.}
\label{fig:layout}
\end{figure}

The conceptual layout of CLIC is shown in the left panel of
Fig.~\ref{fig:layout}~\cite{CLIC}. The basic idea is to
use a relatively high-intensity, low-energy beam to drive a relatively
low-intensity, high-energy beam. The fundamental principle resembles that of a
conventional AC transformer. The low-energy drive beam serves as an RF
source that accelerates the high-energy main beam with a (hopefully) high
accelerating gradient. The left panel of Fig.~\ref{fig:layout} displays the base-line configuration
for a 3-TeV $e^+ e^-$ collider, the primary objective of the CLIC R\&D programme.

Table~\ref{tab:parameters} shows the nominal parameters for CLIC operating at 
its nominal design energy of 3~TeV~\cite{params}. It also shows an alternative set of
parameters for operation at 500~GeV. Note a few key parameters: the nominal
luminosity at each energy is well above $10^{34}$~cm$^{-2}$s$^{-1}$, the main
linac frequency (cf, the 50/60~Hz of a conventional AC circuit)
is now 12~GHz (more similar to the frequency proposed
previously for the NLC and the JLC), the accelerating gradients assumed are
80 (100) MV/m for the 500-GeV (3-TeV) options, and the total site lengths are
13.0 (48.3) km~\cite{CLIC}.

\begin{table}[htbp]
\centering
\epsfig{file=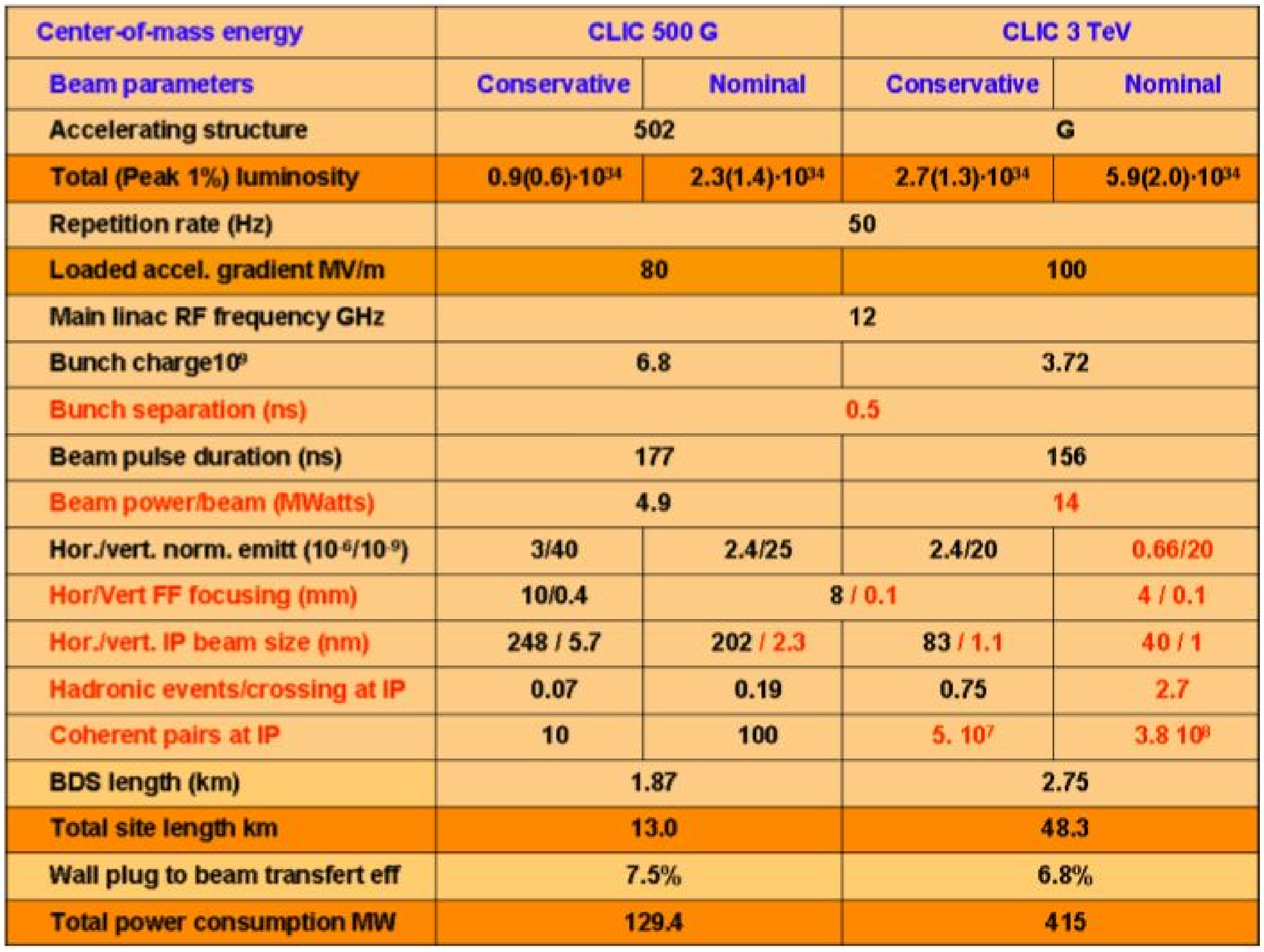,width=3in,angle=90}
\caption{Latest sets of CLIC parameters for $\sqrt{s} = 0.5$~TeV and
the nominal 3~TeV~\protect\cite{params}.}
\label{tab:parameters}
\end{table}

The CLIC technology is less mature than that of the ILC, and requires more R\&D.
In particular, the target accelerating gradient is considerably higher than the ILC, and
requires very aggressive performance from the accelerating structures. The right panel of
Fig.~\ref{fig:layout} shows that the nominal CLIC accelerating gradient
has been exceeded in an unloaded structure
with a very low breakdown probability, below $3.10^{-7}$ per metre, after RF 
conditioning for 1200 hours.
The T18 structure that achieved this performance was designed at CERN,
built at KEK, and RF tested at SLAC. Thus it was the fruit of a truly
international effort.

%
%\begin{figure}[htbp]
%\centering
%{\hspace{1in}
%\epsfig{file=Gradient.ps,width=2in,angle=90}}
%\caption{This is a test figure.}
%\label{fig:gradient}
%\end{figure}
%

The beam gymnastics needed to provide the 12~GHz drive-beam power
source, the RF power generation and two beam acceleration in CLIC standard modules
are being demonstrated in CLIC Test Facility 3, which is being built
by a large international team~\cite{CTF3}.
CLIC R\&D is being carried out by a world-wide collaboration consisting of 24
members representing 27 institutes involving 17 funding agencies from 15 countries
including Turkey.
It is organized in a similar manner to an experimental collaboration, with each of the
institutes represented in a Collaboration Board.

The mandate to the CLIC team is to demonstrate the feasibility of the CLIC
concept by the end of 2010 in a Conceptual Design Report. If this effort is
successful, and if the new physics revealed by the LHC warrants,
the next phase of R\&D on engineering and cost issues could be
completed by the end of 2015. This would serve as the basis for a
Technical Design Report and a request for project approval.

The prospects for approval of the CLIC project would clearly depend not only on its
technical feasibility and cost, but primarily on its physics capabilities and
complementarity to other accelerators such as the ILC. 
These have the subjects of various studies
since 1987, from which the following sections of this talk have been drawn.
The main source is a comprehensive study of CLIC physics published in 2004~\cite{CLICphys},
with significant Turkish participation,
from which a few selected topics are now discussed.

\section*{2 - Light Higgs Physics}

\begin{figure}[htb]
\centering
\epsfig{file=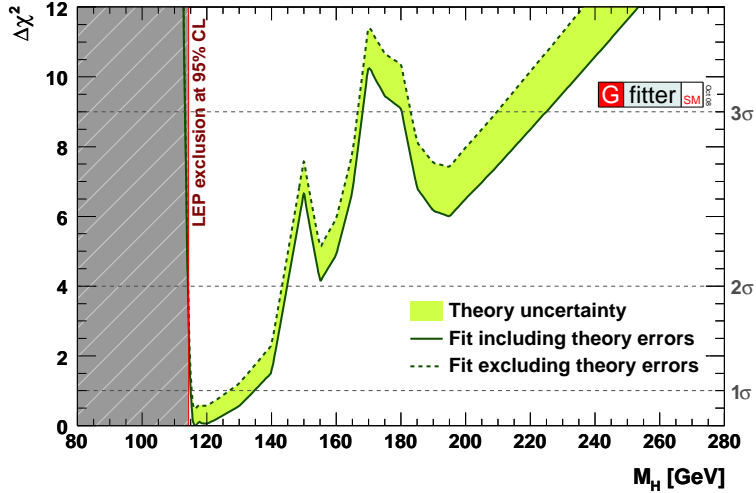,width=4in,angle=0}
\caption{The $\chi^2$ likelihood function for the Standard Model as a function of
the Higgs mass, as obtained~\protect\cite{Gfitter} by combining the indirect information
from the precision electroweak data - which
would yield a smooth, near-parabolic function, and the negative results from the
direct searches at LEP - which cut off low masses, and the Tevatron - which erode
the likelihood between 140 and 200~GeV.}
\label{fig:Higgs}
\end{figure}

We do not yet know whether the Higgs boson exists, still less whether it resembles
the particle predicted in the framework of the Standard Model - and one should
never sell the bearskin before catching the bear! That said, the combined Higgs
probability distribution obtained by combining the direct searches at LEP~\cite{LEP} and
the Tevatron~\cite{Tevatron} with the indirect information provided by high-precision electroweak
measurements~\cite{EWWG} seems to favour a relatively light Higgs boson, as shown in
Fig.~\ref{fig:Higgs}. The electroweak data would, by themselves, yield an almost
parabolic $\chi^2$ function, but this is already being eroded at intermediate values of
$m_h \sim 170$~GeV by the negative results of direct searches at the Tevatron -
and the CDF and D0 searches are continuing. Currently,
the favoured range of Higgs masses is $m_h < 140$~GeV, but masses larger than
200~GeV are still not excluded. Specifically, the Gfitter group finds
\begin{equation}
m_h \; = \; 116.4^{+ 18.3}_{- ~1.3} \; {\rm GeV},
\label{Gfitter}
\end{equation}
and quotes the ranges (114, 145)~GeV at the 68\% confidence level and
(113, 168) and (180, 225)~GeV at the 95\% confidence level~\cite{Gfitter}.

\begin{figure}[htb]
\centering
\epsfig{file=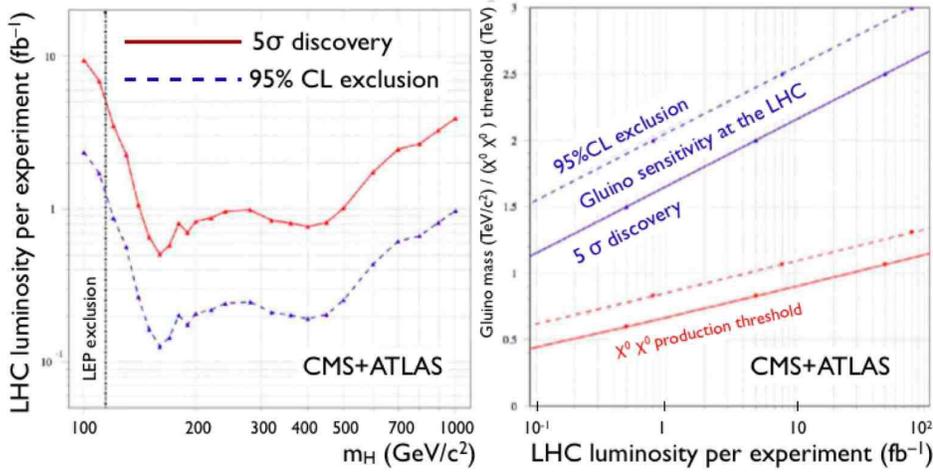,width=5in,angle=0}
\caption{The amounts of integrated LHC luminosity at $E_{CM} = 14$~TeV
required (left) either to exclude a Standard Model Higgs boson at the 95\%
confidence level (blue line) or discover it at the 5-$\sigma$ level (red line), 
and (right) either to exclude a gluino (blue dashed line) or discover it (blue
solid line). The corresponding thresholds for $\chi$ pair production in
$e^+ e^-$ are shown in red~\protect\cite{Blaising}.}
\label{fig:LHCH}
\end{figure}

With just a fraction of 1/fb of integrated luminosity, as seen in 
the left panel of Fig.~\ref{fig:LHCH}~\cite{Blaising},
the ATLAS and CMS experiments would be able to a Standard Model-like Higgs
boson weighing between 140 and 600~GeV. Therefore, either the Tevatron or
the LHC may soon be able to exclude an intermediate-mass Higgs and tell us that
it must either be very light, close to the LEP lower limit, or else very heavy. What
could CLIC contribute in either of these scenarios?

\begin{figure}[htb]
\centering
{\epsfig{file=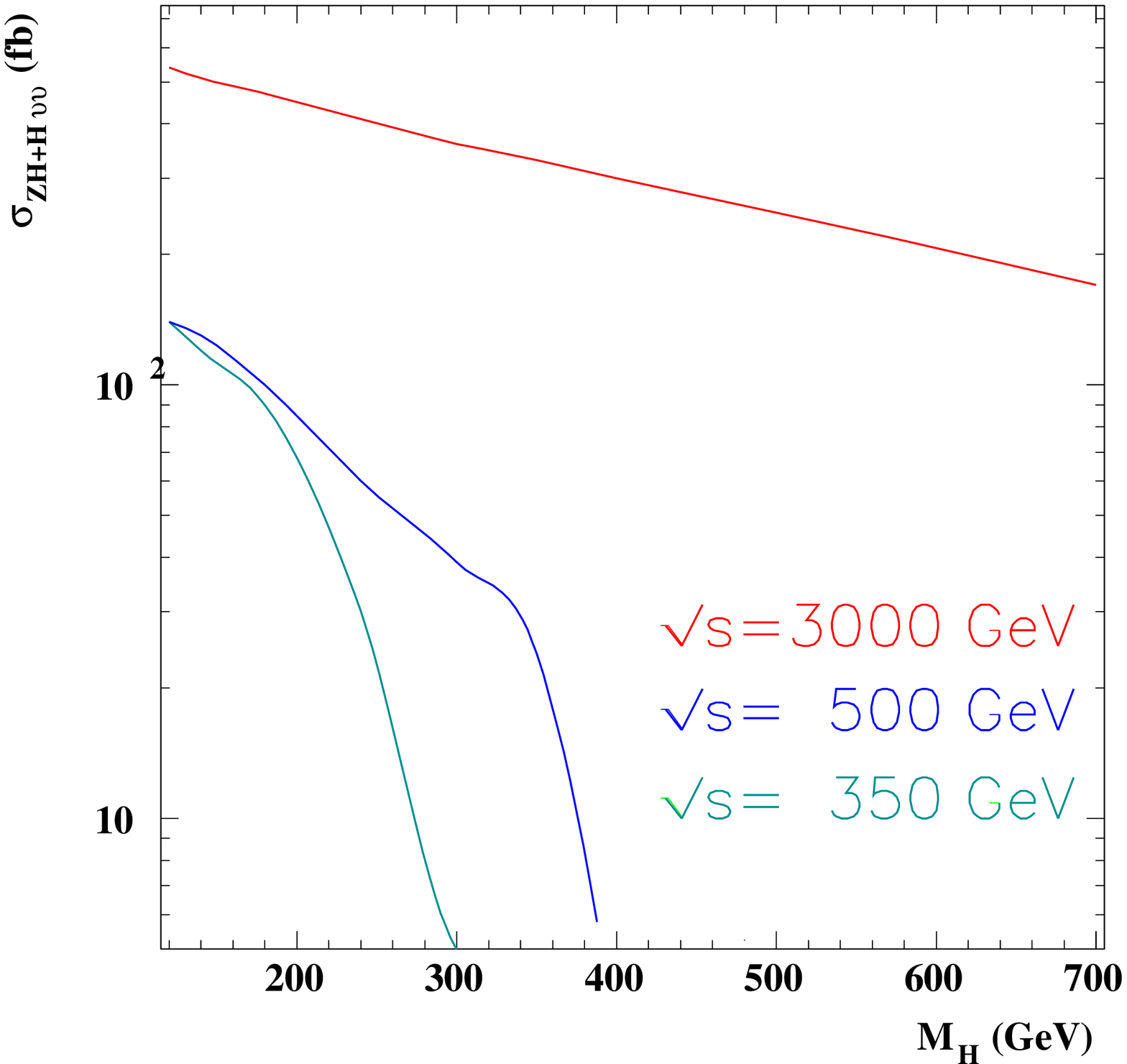,width=2.55in,angle=0}
\epsfig{file=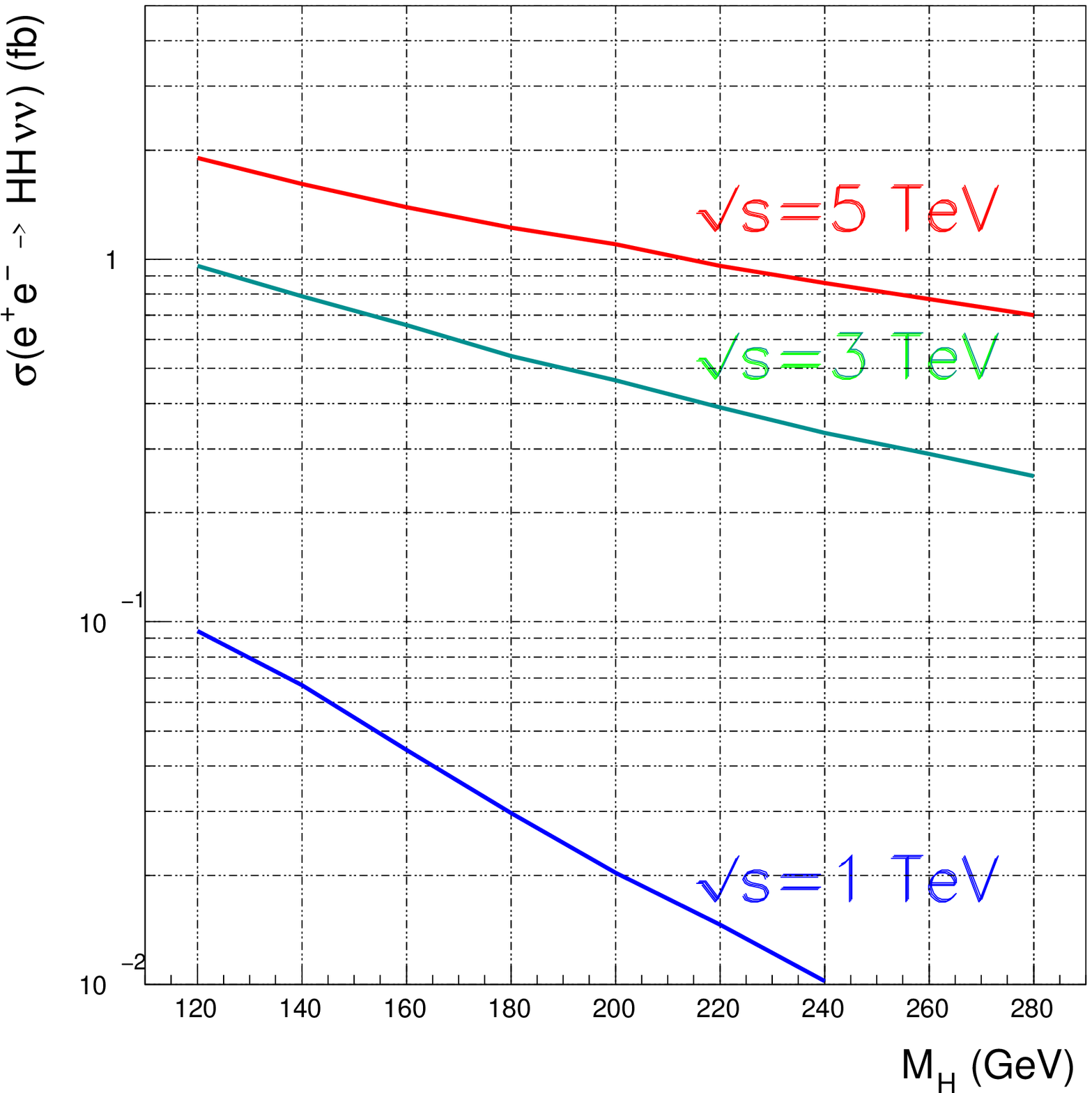,width=2.45in,angle=0}}
\caption{Inclusive single Higgs production cross section (left) and double Higgs production
(right) as functions of the Higgs mass, each for three values of the $e^+e^-$ centre-of-mass
energy~\protect\cite{CLICphys}.}
\label{fig:HXsection}
\end{figure}

\begin{figure}[htb]
\centering
{\epsfig{file=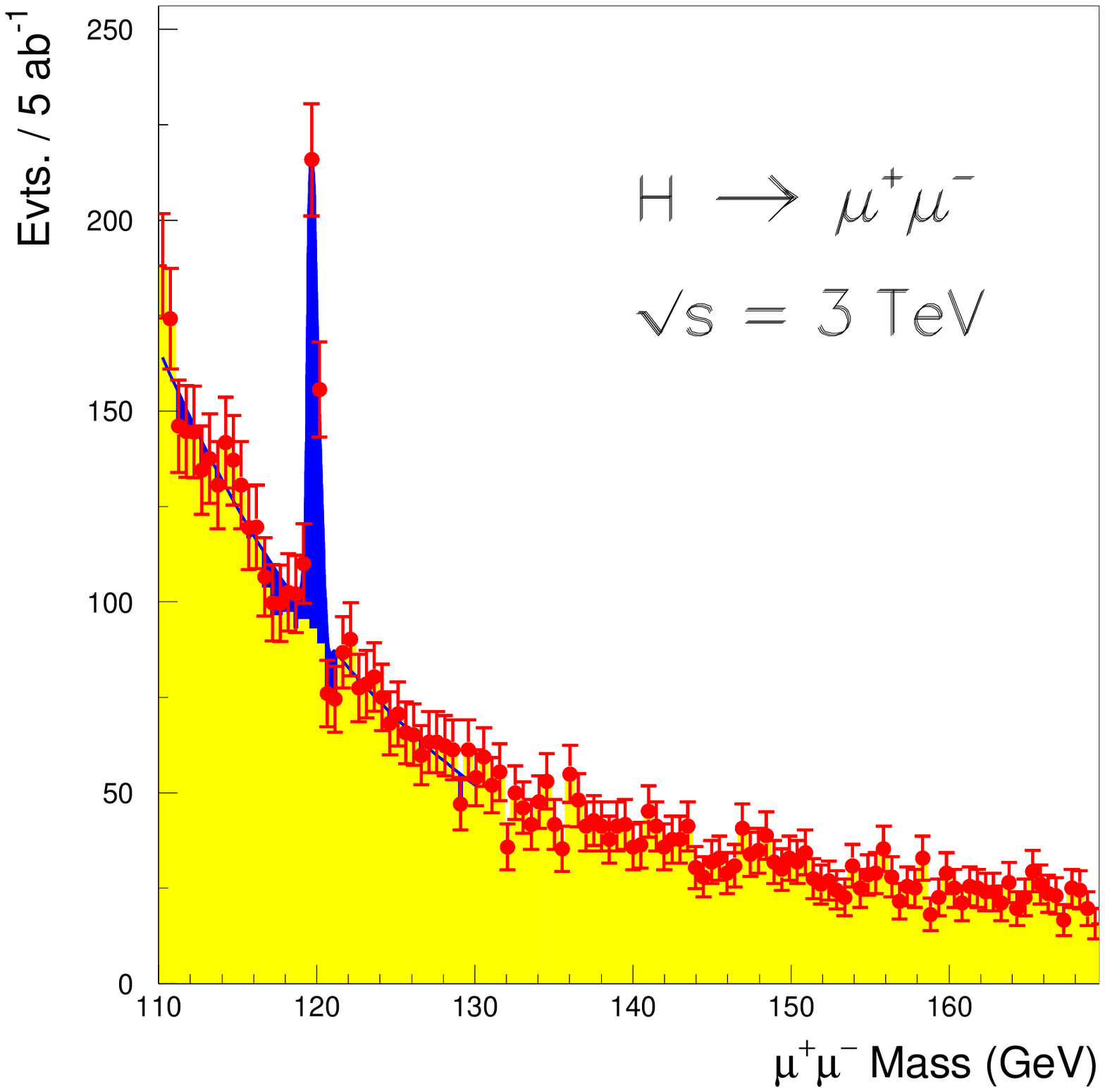,width=2.5in,angle=0}
\epsfig{file=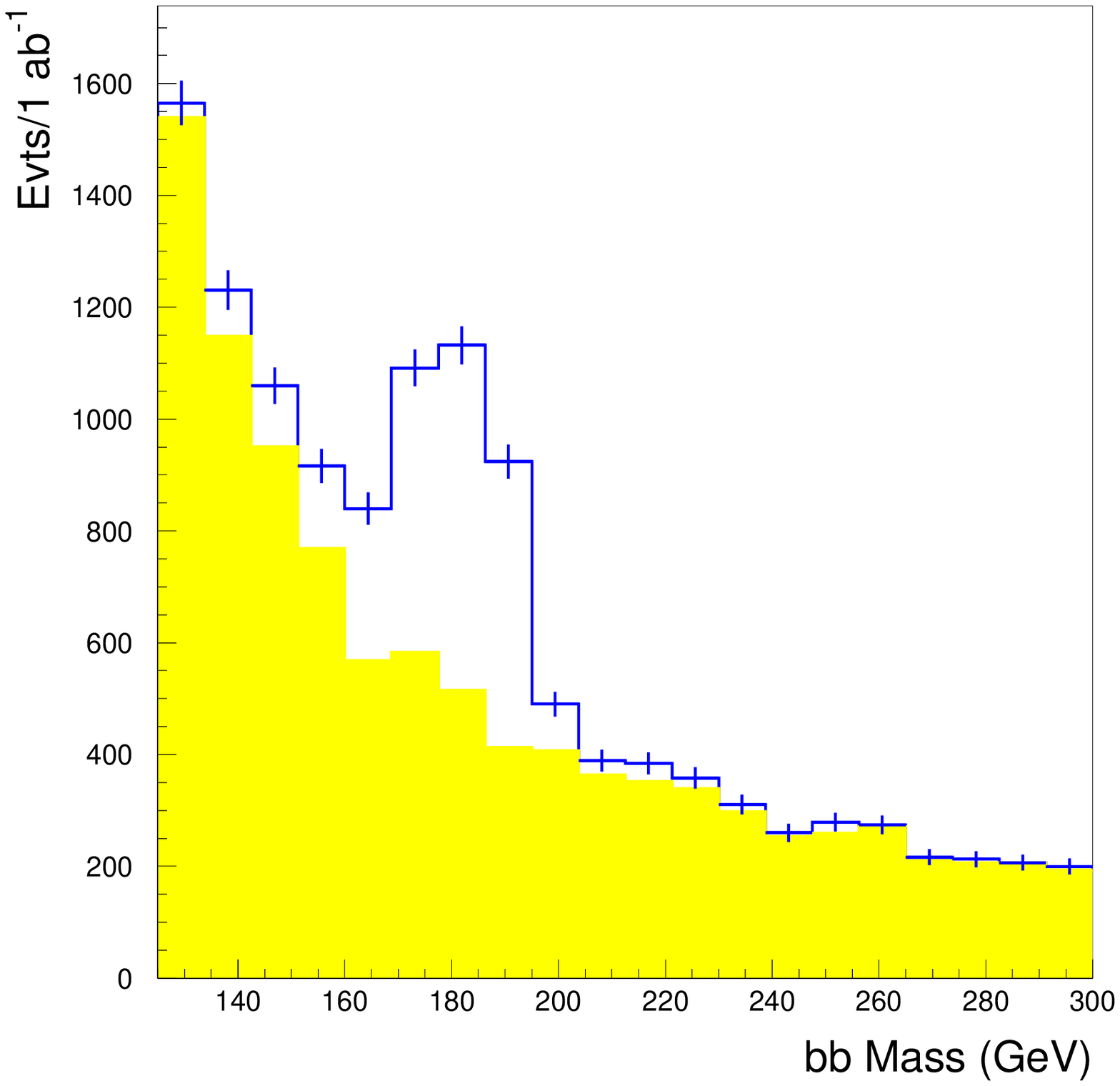,width=2.5in,angle=0}}
\caption{Reconstructed signals for $H \to \mu^+\mu^-$ (left) and
$H \to b\bar{b}$ (right) for $M_H = 120$~GeV and 180~GeV,
respectively, at $\sqrt{s}$~=~3~TeV~\protect\cite{CLICphys}.}
\label{fig:rare}
\end{figure}

If there is a light Higgs boson, the ILC will be able to study many of its
properties in some detail. The cross section for producing it at CLIC will
be even much larger than at the ILC, as seen in the left panel of Fig.~\ref{fig:HXsection}. 
The increase compared with lower centre-of-mass energies is more pronounced for higher
$m_h$, but is substantial already for $m_h \sim 120$~GeV.
This increase will open up the possibility of measuring rare Higgs decays which are
unobservable at the LHC and difficult to measure at a lower-energy $e^+ e^-$
collider, and two examples are displayed in Fig.~\ref{fig:rare}. In the left panel we
see that CLIC could measure the $h \mu^+ \mu^-$ coupling with an accuracy of 4\%
if $m_h = 120$~GeV, and in the right panel we see that CLIC could measure the 
$h {\bar b} b$ coupling with an accuracy of 2\% if $m_h = 180$~GeV~\cite{CLICphys}.


\begin{figure}[htb]
\centering
{\epsfig{file=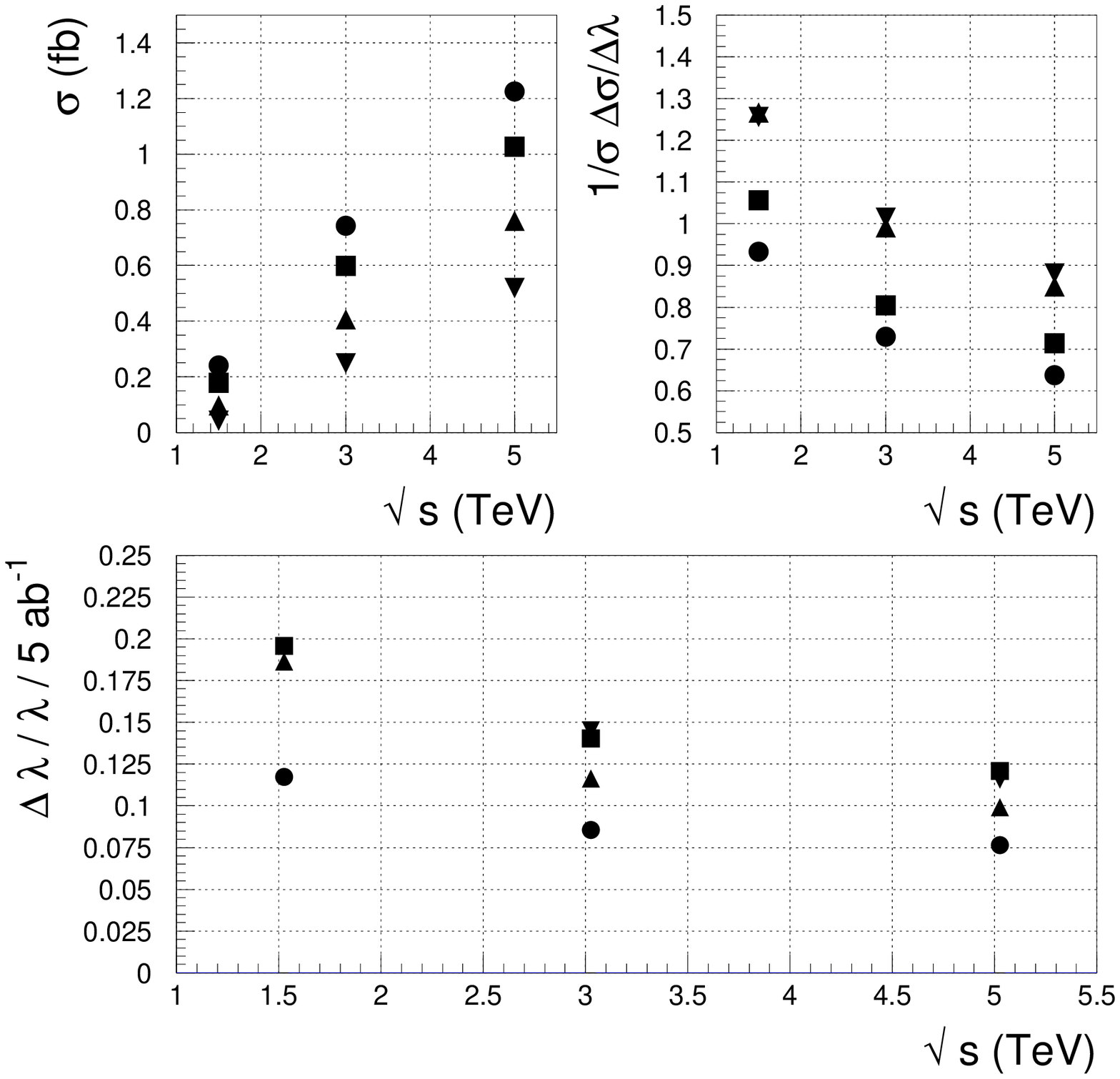,width=2.25in,angle=0}
\epsfig{file=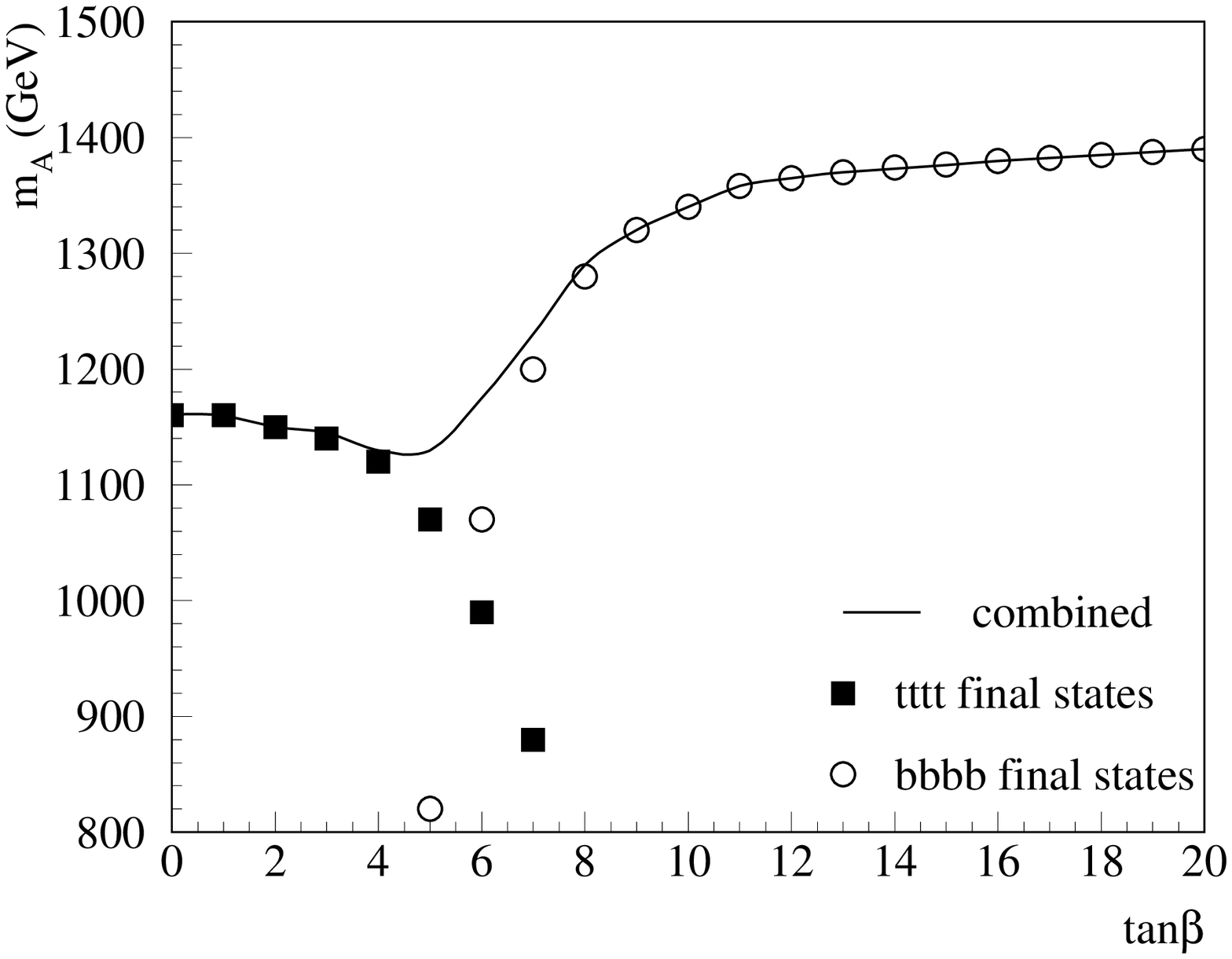,width=2.75in,angle=0}}
\caption{Left: The cross section of the process $e^+e^- \rightarrow H\nu\nu$, 
the sensitivity of the cross section to the triple 
Higgs coupling, and the expected precision with which  the triple
Higgs coupling could be measured, for $m_h = 120$~GeV (circles), 140~GeV (squares), 
180~GeV (triangles), and 240~GeV (inverted triangles), assuming 5/ab of integrated
luminosity. Right: The $H A$ discovery reach with 3/ab of CLIC data 
at $\sqrt{s} = 3$~TeV as a function of ${\rm tan}\beta$ obtained by summing the
$b \bar {b} \bar{b} b$ and $t \bar{t} \bar{t} t$ channels~\protect\cite{CLICphys}.}
\label{fig:dellambda}
\end{figure}

The double-Higgs production cross section at CLIC would also be much larger
than at lower energies, as seen in the right panel of Fig.~\ref{fig:HXsection}~\cite{CLICphys}. 
As a result, if the Higgs 
mass is in the low-mass region, the triple-Higgs coupling could be measured quite 
accurately at CLIC: to 11\% if $m_h = 180$~GeV, or to 9\% if $m_h = 120$~GeV,
as seen in the left panel of Fig.~\ref{fig:dellambda}. Because of the higher cross sections at
higher centre-of-mass energies, the measurement at CLIC could be significantly
more accurate than at a lower-energy $e^+ e^-$ collider.

\section*{3 - Accompanying New Physics?}

If $m_h$ is as light as 120~GeV, the present electroweak vacuum is rendered
unstable by radiative corrections induced by the top quark, unless new physics
intervenes. One possibility for this is some
form of contact interaction, and the high CLIC centre-of-mass energy gives it an
edge over a lower-energy collider in searching for such a symptom of new
physics. Studies show that
CLIC would be sensitive to new contact interactions in $e^+ e^- \to \mu^+ \mu^-$
with scales up to 300~TeV~\cite{CLICphys}.

One of the most compelling examples of possible new physics is supersymmetry,
which would help stabilize the electroweak vacuum~\cite{ER}. Supersymmetry
is discussed later in its own right. Here I note that one of its predictions is the
existence of heavier neutral Higgs bosons, a pseudoscalar $A$ and a scalar $H$,
which would be quite difficult to detect at the LHC, depending on their masses. 
In general, an $e^+ e^-$ collider 
could extend the search up to close to the kinematic limit. This sensitivity is
exemplified for CLIC in the right panel of Fig.~\ref{fig:dellambda}, 
where we see that a pseudoscalar $A$
boson would be detectable at CLIC if its mass were up to 1100~GeV or more,
depending on the value of $\tan \beta$, the ratio of Higgs VEVs in the minimal
supersymmetric extension of the Standard Model (MSSM)~\cite{CLICphys}.

\section*{4 - Theorists getting Cold Feet?}

With the imminent discovery (or demise) of the Higgs boson, many theorists
seem to be getting cold feet: can it really be true? Now may be the last chance
to stake a claim to an alternative theory, and many theorists are seizing it.

Maybe the Higgs boson does exist, and is even light, as in little Higgs models~\cite{little}?
In these models, the Higgs is a pseudo-Nambu-Goldstone boson that is a bound
state of new strong-interaction dynamics that appears at $\sim 10$~TeV.
One-lop quadratic divergences in the Higgs mass-squared are cancelled by an
extra `top-likeÕ quark, gauge bosons and extra scalars related to the Higgs boson,
all with masses $\sim 1$~TeV. These would be prime fodder for discovery at CLIC.
 
Alternatively, perhaps our interpretation of the high-precision electroweak data
is incorrect~\cite{Chan}? As is well-known, there are some apparent discrepancies between
different subsets of the electroweak data.  For example, measurements of
hadronic final states in $Z$ decays tend to prefer a higher value of $m_h$ than
do measurements of leptonic final states. Also, the low-energy measurement of
$\sin^2 \theta_W$ by the NuTeV collaboration seems to differ from other
measurements~\cite{Chan}. Most observers would consider these discrepancies to be
symptomatic of underestimated systematic errors, or some other experimental
problems. But perhaps they are due to some unknown new physics? In that
case, the Standard Model would be an incomplete paradigm for analyzing the
high-precision electroweak data, the apparent preference for a low-mass Higgs
boson might be wrong. In that case, a there might be a heavier Higgs boson,
ripe for observation at CLIC.

\begin{figure}[htb]
\centering
\epsfig{file=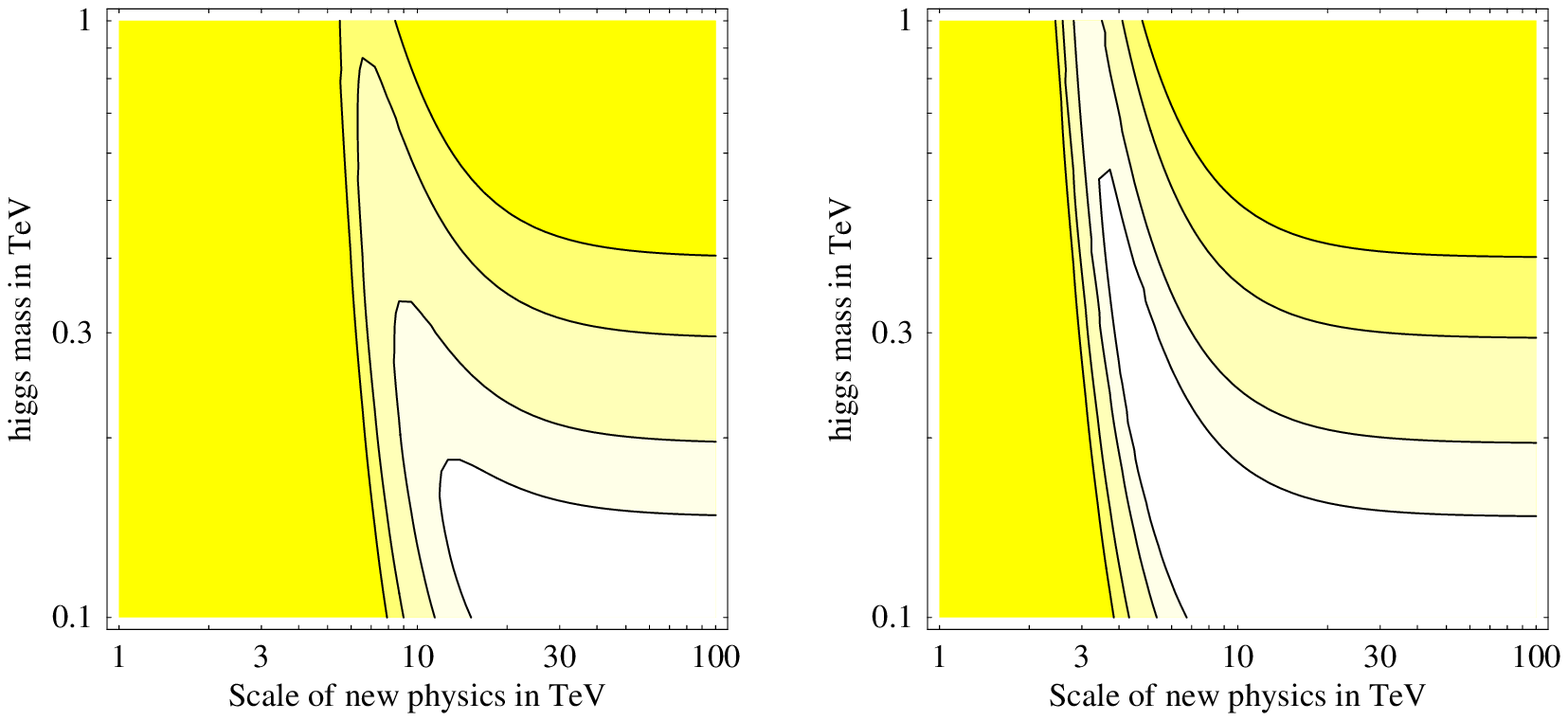,width=5in,angle=0}
\caption{Fits to the electroweak data that include one or another higher-dimensional
operator scaled by a high scale $\Lambda \sim 3 - 10$~TeV
may allow a heavier Higgs boson~\protect\cite{BS}.}
\label{fig:corridor}
\end{figure}

Another corridor towards a heavier Higgs boson might be opened up by the
inclusion in the electroweak fit of higher-dimensional operators composed of 
Standard Model fields. As shown in Fig.~\ref{fig:corridor}, if one such operator
is present with a coefficient scaled by a high scale $\Lambda \sim 3 - 10$~TeV
(possibly generated by the exchange of some massive state), a Higgs weighing
up to 1~TeV might be compatible with the high-precision electroweak data~\cite{BS}.

One example of a theory with massive states that invalidate the electroweak
`bound' is provided by a model with a fourth generation~\cite{ACS}. In this model, 
the Standard Model
electroweak fit could accommodate (even prefer) a heavier Higgs weighing
$\sim 300$~GeV~\cite{Kribs}, for which the cross section at CLIC would be
encouragingly large, as seen in Fig.~\ref{fig:HXsection}.

Finally, let us mention Higgsless models~\cite{Hless}. These are beset by strong $WW$ 
scattering, which tends to feed back into an unacceptable fit to the
high-precision electroweak data. This problem can be somewhat mitigated 
in variants with extra dimensions, but is still a serious issue for such models.

\section*{5 - What if the Higgs is Heavy - or Non-Existent?}

\begin{figure}[htb]
\centering
\epsfig{file=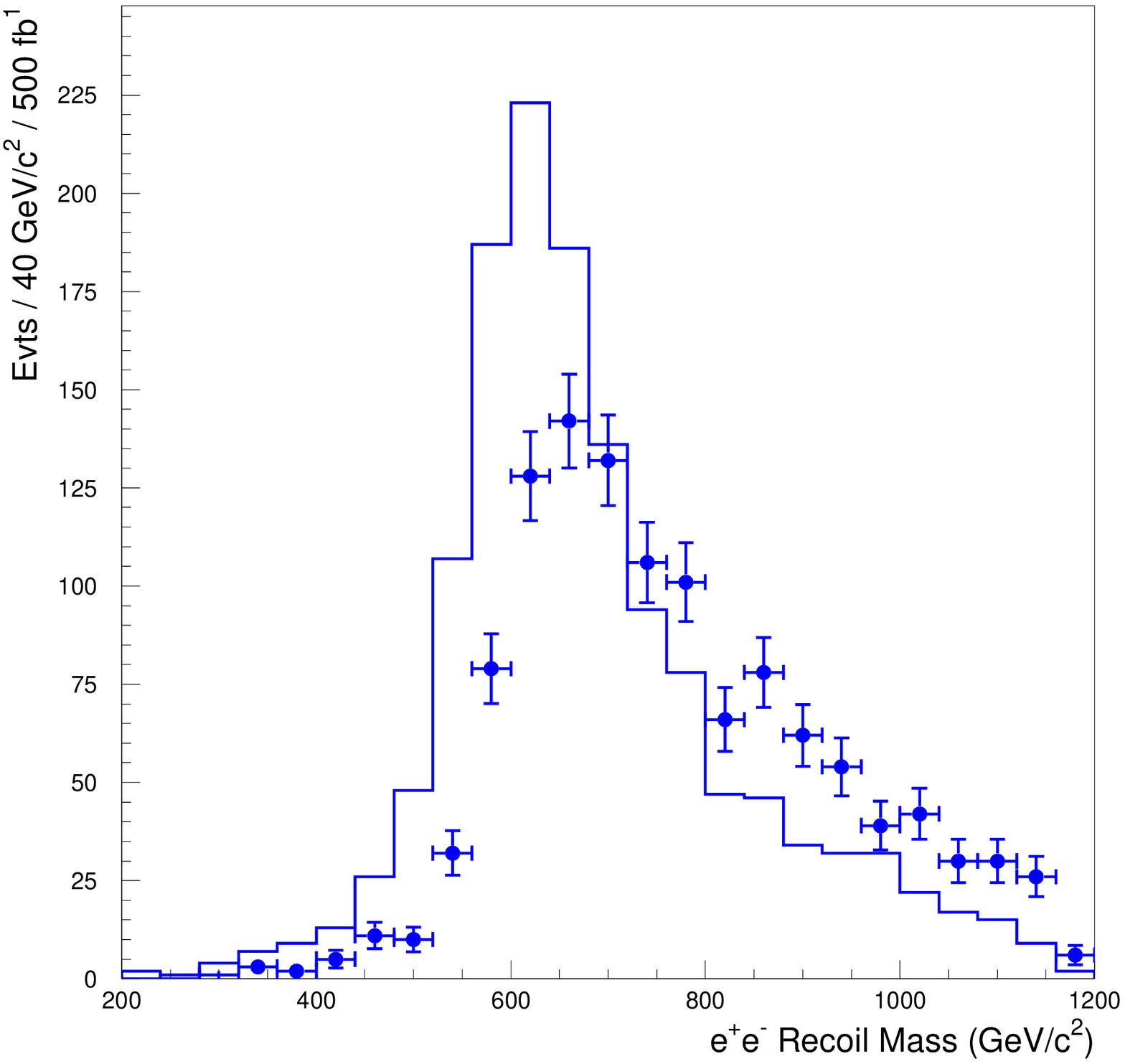,width=2.5in,angle=0}
\caption{The invariant mass recoiling against an $e^+ e^-$ pair at CLIC
operating at a nominal $\sqrt{s} = 3$~TeV in the case of a heavy Higgs 
boson~\protect\cite{CLICphys}.}
\label{fig:HeavyH}
\end{figure}

If the Higgs boson does indeed weigh 1~TeV or so, its observation will be difficult
at the LHC, though not impossible, thanks to the $WW$ fusion
mechanism for Higgs production. However, such a heavy Higgs boson would
not be visible directly at the ILC. There would be no problem producing and
measuring it at CLIC, as seen in Fig.~\ref{fig:HeavyH}, which shows
the recoil mass distribution for a heavy Higgs boson produced in the reaction $e^+ e^- \to 
e^+ e^- H$, in a simulation for a nominal $M_H = 900$~GeV~\cite{CLICphys}.

If there is no Higgs boson at all, the LHC might find a hint of strong $WW$
scattering, but this new physics would not be visible directly at a lower-energy 
$e^+ e^-$ collider. On the other hand, either the LHC or the ILC might be
able to discover associated physics related, e.g., to extra dimensions.
However, CLIC would be uniquely well placed to study strong $WW$ scattering
directly, with high statistics and precision. CLIC would also be best placed to 
see/understand other aspects of scenarios with associated high-scale
physics, such extra dimensions or composite models of Higgs, quarks
and leptons.

\section*{6 - Supersymmetry}

There are many reasons to like supersymmetry, including its intrinsic beauty,
its utility in controlling radiative corrections to the Higgs mass and thereby
solving the naturalness aspect of the hierarchy problem~\cite{hierarchy}, the help it offers for
the unification of the gauge couplings~{GUTs}, the fact that it predicts a light Higgs
boson weighing $< 150$~GeV~\cite{erz}, as apparently favoured by the precision 
electroweak data~\cite{EWWG,Gfitter}, and the fact that it predicts naturally the existence and a
suitable density for the astrophysical cold dark matter~\cite{EHNOS}. Moreover,
supersymmetry is an (almost) essential ingredient in string theory.

These are all good arguments but, to paraphrase Feynman, if we
had one {\it really} convincing argument, we would not need to give several!

\begin{figure}[htb]
\centering
{\epsfig{file=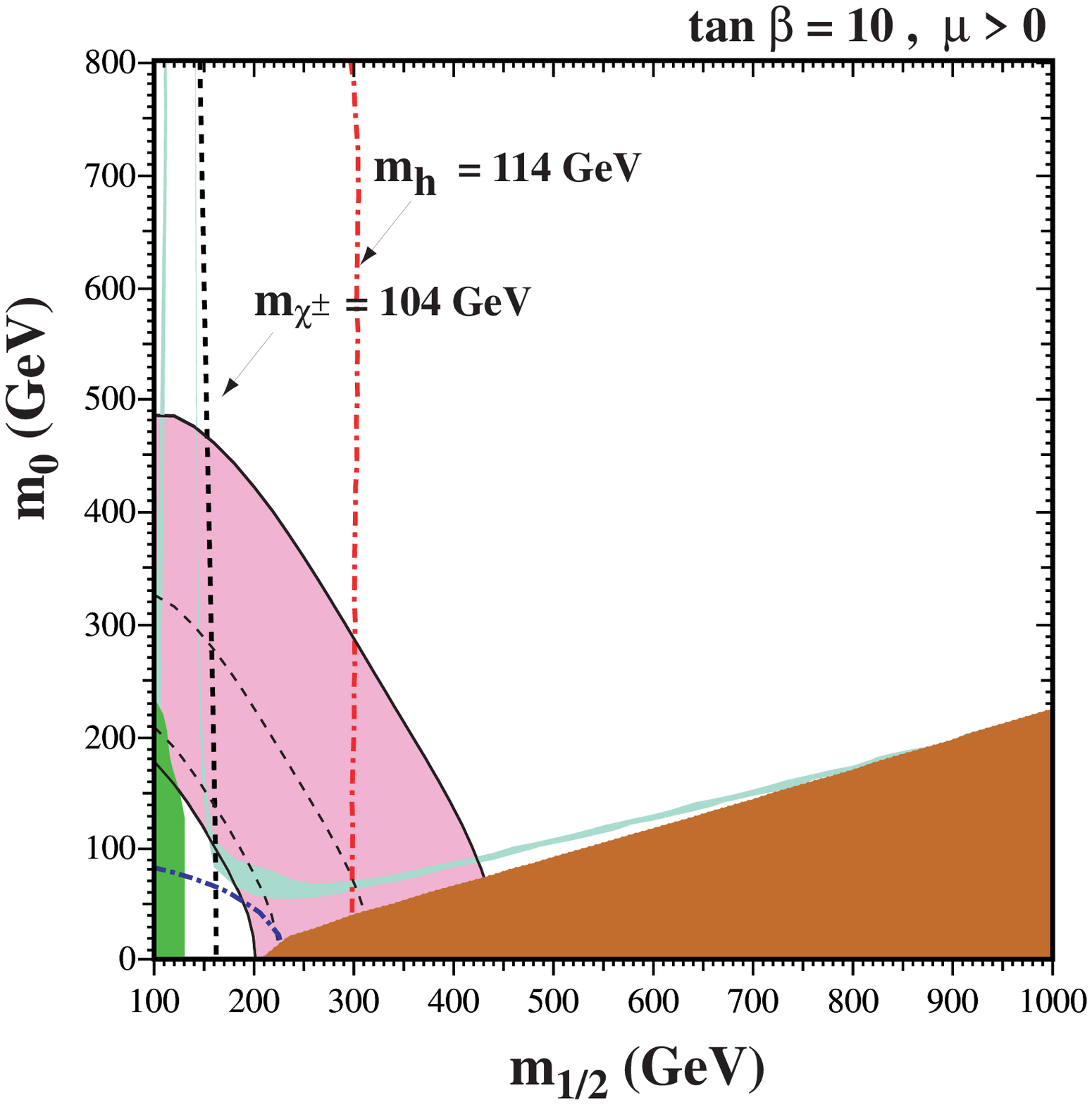,width=2.25in,angle=0}
\epsfig{file=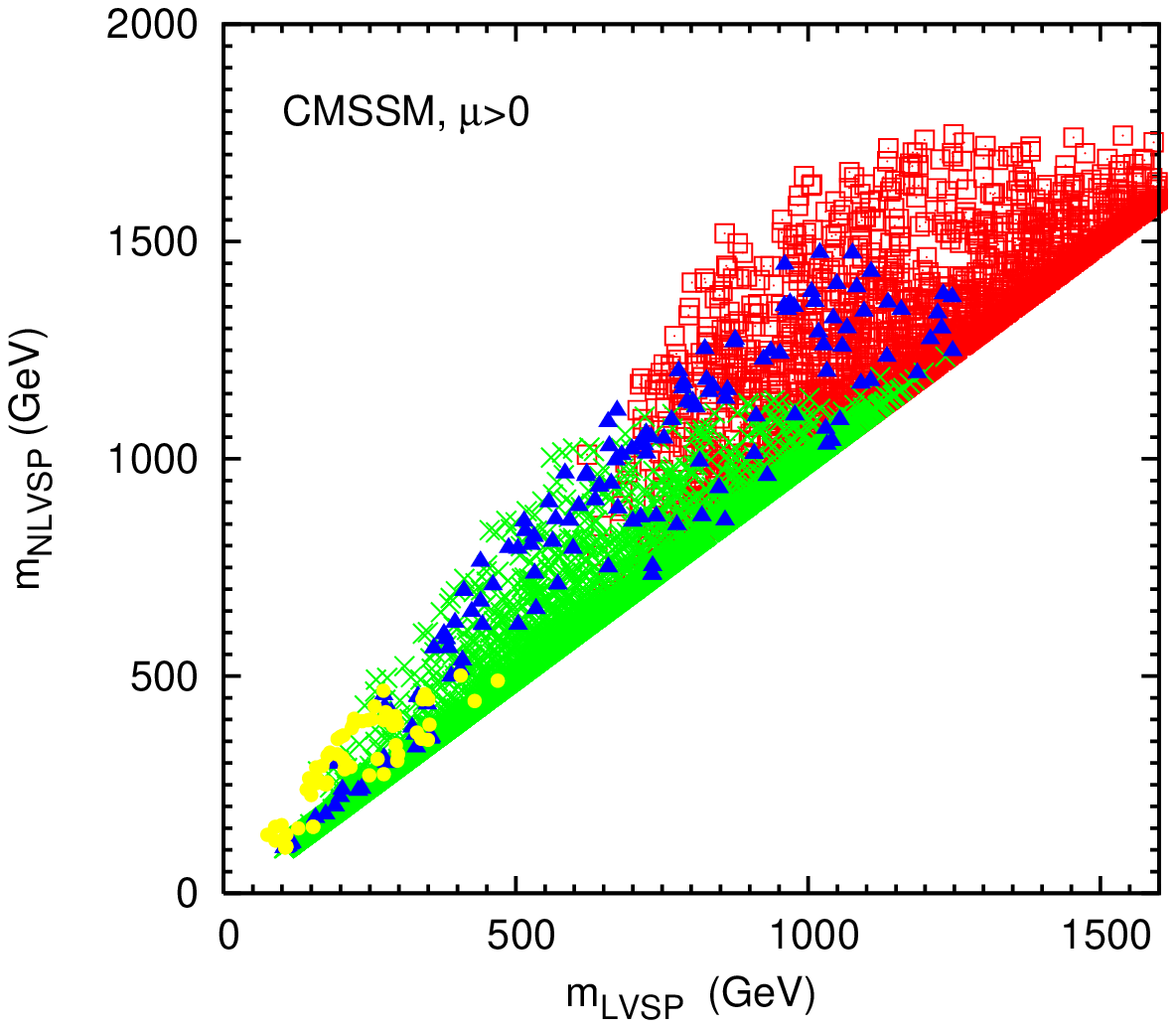,width=2.75in,angle=0}}
\caption{Left: The $(m_{1/2}, m_0)$ plane in the CMSSM for $\tan \beta = 10$,
$A_0 = 0$ and $\mu >0$, showing the impacts of different phenomenological
constraints explained in the text~\protect\cite{EOSS}. 
Right: Scatter plot of the masses of the lightest visible sparticle 
(LVSP) and next-to-lightest visible sparticle (NLVSP) in the CMSSM: the red open squares
represent the full model sample, the blue triangles the points providing a suitable
density of cold dark matter, the green crosses the points accessible to the LHC, and the
yellow circles those amenable to direct dark matter detection~\protect\cite{LVSP}.}
\label{fig:susy}
\end{figure}

The left panel of Fig.~\ref{fig:susy} compiles the constraints on the simplest version of the
minimal supersymmetric extension of the Standard Model, in which
the scalar and spin-1/2 sparticle masses are each constrained to be all equal
to $m_0$ and $m_{1/2}$, respectively,
at the grand unification scale (the CMSSM), assuming that the lightest
supersymmetric particle (LSP) is the lightest neutralino (a mixture of partners 
of the photon $Z$ and Higgs boson)~\cite{EOSS}. The bottom-right part of the $(m_{1/2}, m_0)$
plane is excluded in this case, because there the LSP is the charged stau.
Regions at low $m_{1/2}$ are excluded by LEP searches for charginos and
the Higgs boson, and by $b \to s \gamma$ decay. There is also a diagonal (pink) band
that is favoured if one uses the published low-energy $e^+ e^-$ data to calculate 
$g_\mu - 2$~\cite{g-2,g-2th}. However, this is a controversial constraint, so we will treat it
as optional. Finally, note the thin diagonal (turquoise) strip within which the relic LSP
density matches that inferred from astrophysics and cosmology by WMAP and
other experiments. Combining all these constraints, we see that there is a limited
region of the WMAP strip that is compatible with all the constraints, but that this
extends to relatively large values of $m_{1/2}$ and hence sparticle masses,
if the potential $g_\mu - 2$ constraint is discounted.

The gluino mass increases proportional to $m_{1/2}$
along this WMAP strip, and the lightest neutralino
is also simply proportional. Therefore, a given reach in the gluino mass
translates directly into an LSP mass, and hence a threshold for sparticle
production in $e^+ e^-$ collisions, as seen in the right panel of Fig.~\ref{fig:LHCH}~\cite{Blaising}. 
With just 100/pb of data at 14~TeV, the LHC
should be able to discover the gluino if it weighs less than 1.1~TeV, or exclude a
gluino weighing less than 1.5~TeV if it sees nothing. In the former case, the 
threshold for sparticle pair production in $e^+ e^-$ collisions would be below
0.5~TeV, and hence accessible to the ILC. However, in the latter case, the
sparticle threshold would necessarily lie above 0.6~TeV. More generally, at
least in simple supersymmetric models, the LHC will tell $e^+ e^-$ colliders
what energy they need to observe supersymmetry.

The extension of the WMAP strip to high mass scales in the left panel of Fig.~\ref{fig:susy}
shows that sparticles might be quite heavy. As seen in the right panel of Fig.~\ref{fig:susy},
the LHC would be able to discover supersymmetry in most (but not all) of
the parameter space of the CMSSM. An $e^+ e^-$ centre-of-mass energy
of 1~TeV would cover only a part of the WMAP-compatible parameter
space, whereas 3~TeV should be enough to cover (almost) all of it, 
at least in the CMSSM~\cite{LVSP}.

\section*{7 - How Soon might Supersymmetry be Detected?}

So far we have been treating all the constraints on supersymmetric
models as if they were $\theta$-functions. What happens if one makes a
frequentist likelihood analysis, taking the $g_\mu - 2$ indication at its
face value?

\begin{figure}[htb]
\centering
\epsfig{file=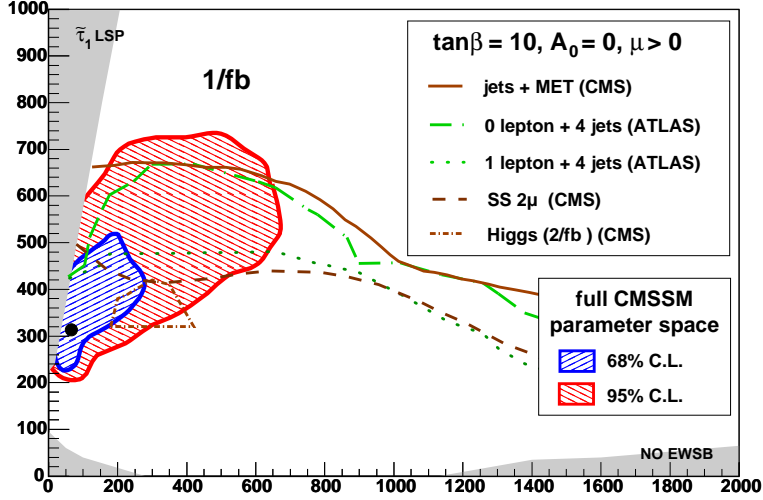,width=4in,angle=0}
\caption{The $(m_0, m_{1/2})$ plane in the CMSSM:
the best-fit point is indicated by a filled circle, and the 
  68 (95)\% confidence-level contours from our fit are shown as dark grey/blue (light
  grey/red) overlays, scanned over all $\tan\beta$ and $A_0$ values~\protect\cite{Master}.
Also shown are some 5-$\sigma$ discovery contours at ATLAS~\protect\cite{ATLAS}
and CMS~\protect\cite{CMS} with
  1~fb$^{-1}$ at 14~TeV, and the contour for the $5\,\sigma$ discovery
  of the Higgs boson in sparticle decays with 2~fb$^{-1}$ at 14~TeV in CMS.
  These were estimated assuming $\tan\beta = 10$ (similar to our best-fit value)
  and $A_0 = 0$.}
\label{fig:Master1}
\end{figure}

Fig.~\ref{fig:Master1} displays the 68\% and 95\% confidence level regions
in the $(m_0, m_{1/2})$ plane for the CMSSM (similar results are found if one
relaxes scalar-mass universality for the Higgs multiplets)~\cite{Master}. We see that much
of the 68\% region would be covered already with 50/pb of integrated
luminosity with the LHC at 10~TeV, and all of it with 100/pb at 14~TeV.
We also see that supersymmetry would be discovered by the LHC with 1/fb
over almost all the 95\% region, and this amount of integrated luminosity
would also suffice to exclude supersymmetry throughout the 95\% region.

\begin{figure}[htb]
\centering
{\epsfig{file=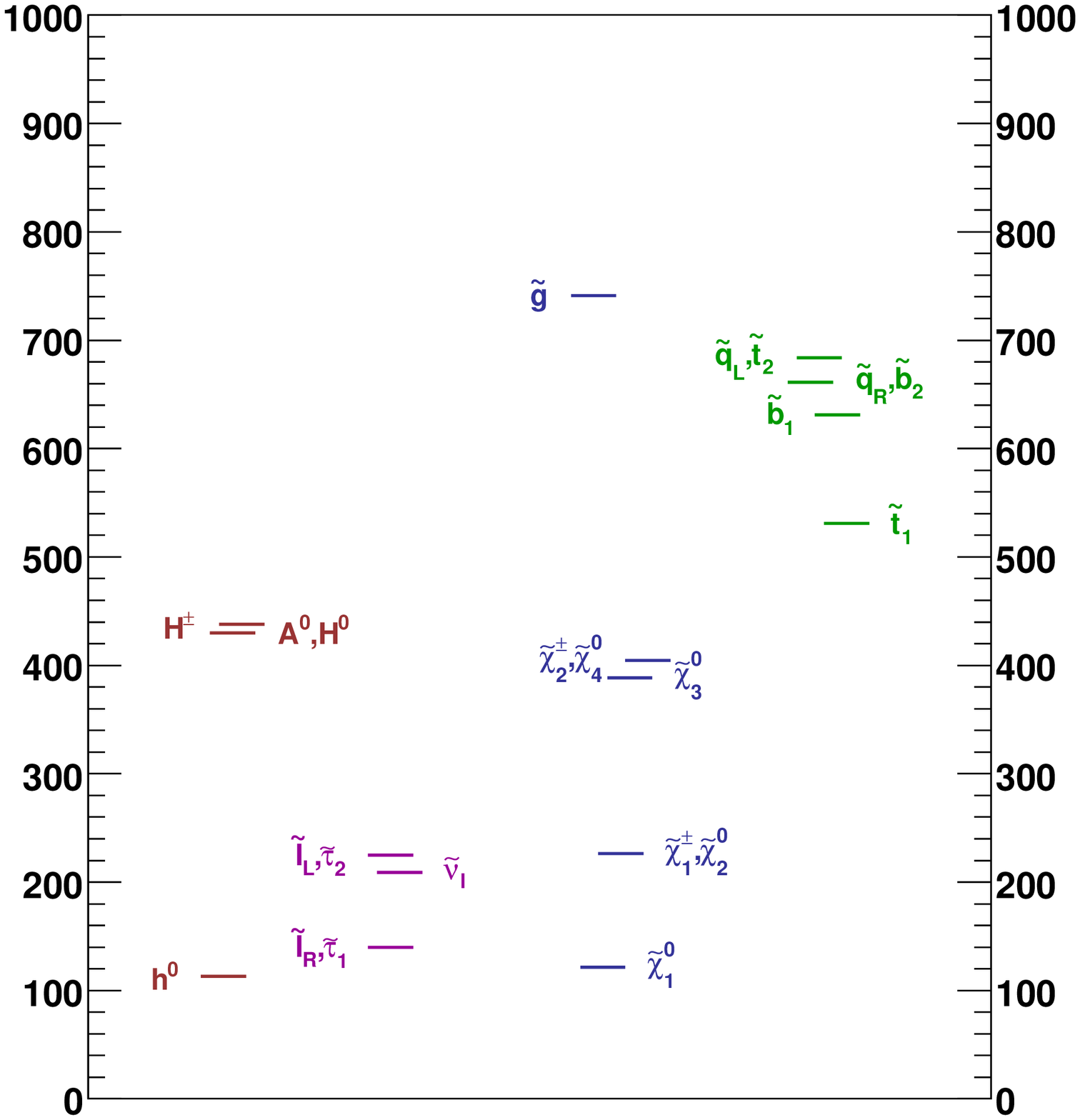,width=2.2in,angle=0}
\epsfig{file=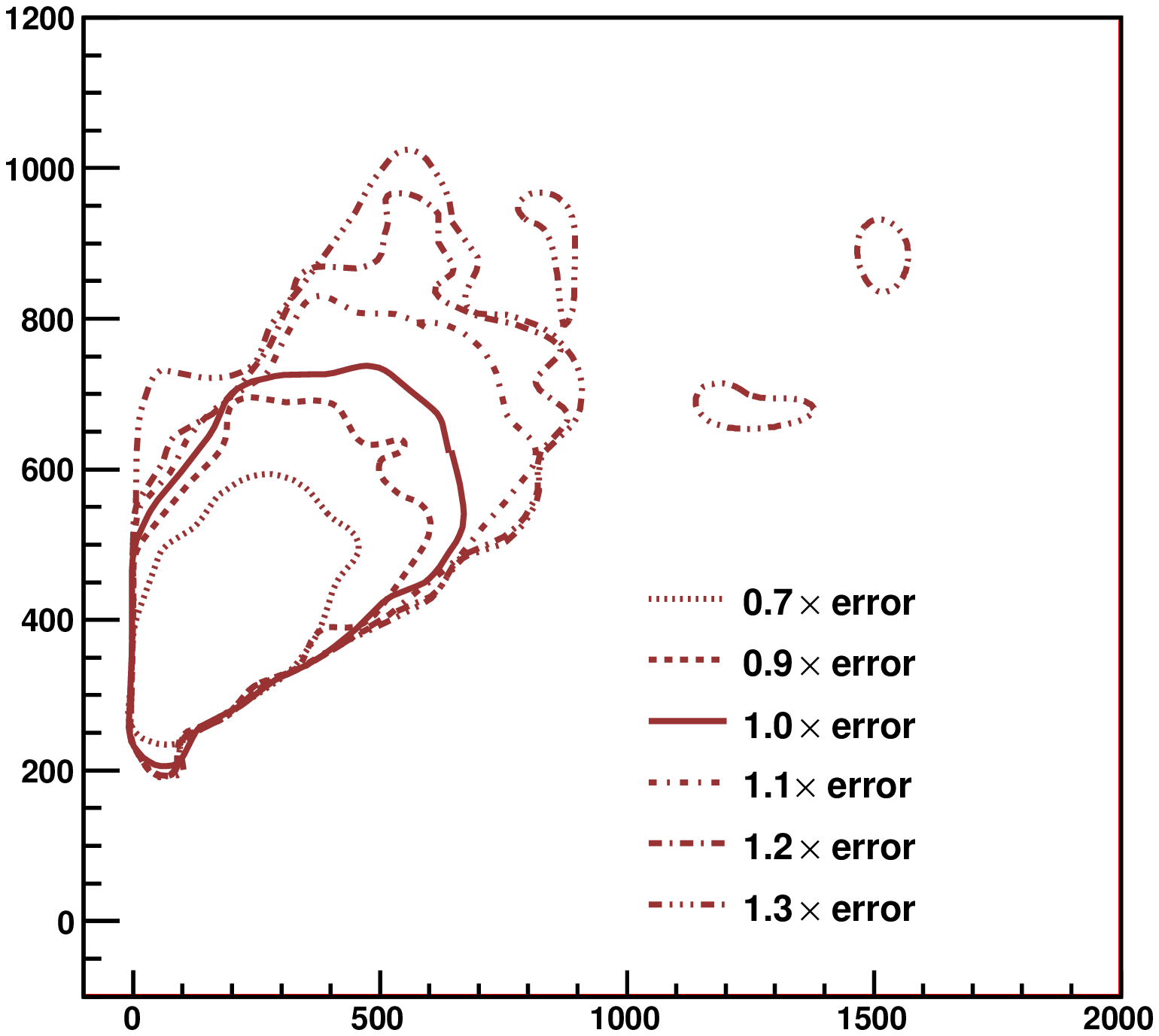,width=2.8in,angle=0}}
\caption{Left: The spectrum at the best-fit CMSSM point shown in 
Fig.~\protect\ref{fig:Master1}. Right: The variation of the preferred region in the
CMSSM $(m_0, m_{1/2})$ plane as the error in the theory-experiment comparison
for $g_\mu - 2$ is rescaled~\protect\cite{Master}.}
\label{fig:spectrum}
\end{figure}

The best-fit point in the CMSSM has $\tan \beta \sim 10$ and quite low
$m_{1/2}$, similar to benchmark point B~\cite{Bench} or SPS1a~\cite{SPS}. As such, it has a relatively
light spectrum, that offers good opportunities to the ILC. The best-fit spectrum
is shown in the left panel of Fig.~\ref{fig:spectrum}, where we see that an $e^+ e^-$ collider with 
0.5~TeV could produce all the sleptons, the lighter chargino and the second-lightest
neutralino. One would need $\sim 1$~TeV to produce the heavier neutral and
charged Higgs bosons, and to pair-produce the heavier charginos and
neutralinos. A centre-of-mass energy above 1~TeV would be needed to
produce squarks. Thus, even in this encouraging example, there would be work
for both the ILC and CLIC. This is just one illustration that, even in low-mass
supersymmetric scenarios where the ILC can produce some sparticles, 
studies of strongly-interacting sparticles would require the higher
centre-of-mass energy of CLIC.

An important word of warning: the result of this likelihood analysis depends
sensitively on the treatment of $g_\mu - 2$. If one rescales the error in the
comparison between theory and experiment, as seen in the right panel of Fig.~\ref{fig:spectrum},
the preferred region of the $(m_0, m_{1/2})$ plane expands and contracts
considerably~\cite{Master}. (The same is true for rescaling the error in $b \to s \gamma$.)
Moreover, if one uses $\tau$ decay data to calculate $g_\mu - 2$ in the Standard Model,
the discrepancy with experiment essentially disappears, and very large
sparticle masses beyond the reach of the ILC are allowed, even favoured.

\begin{figure}[htb]
\centering
{\epsfig{file=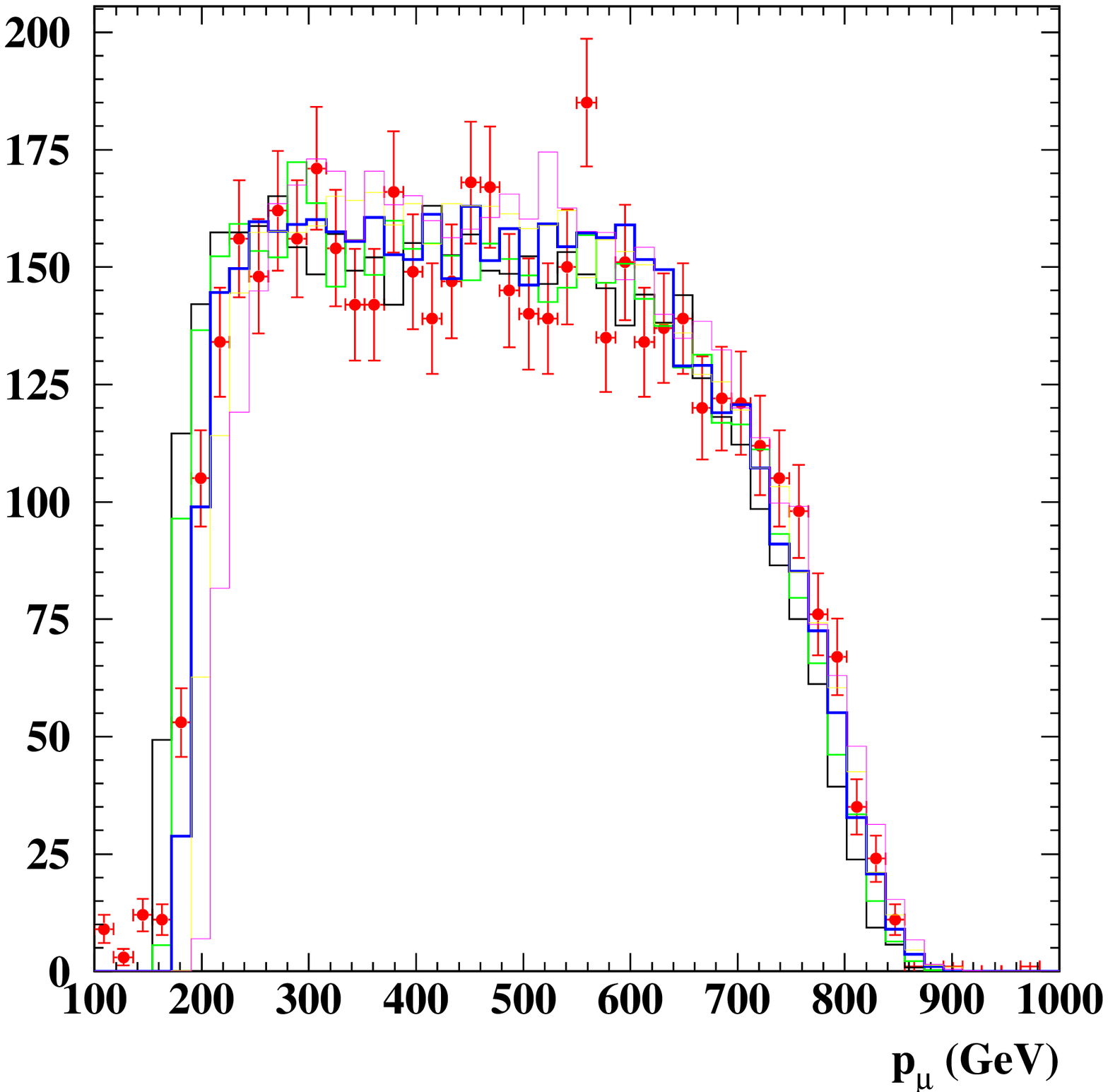,width=2.5in,angle=0}
\epsfig{file=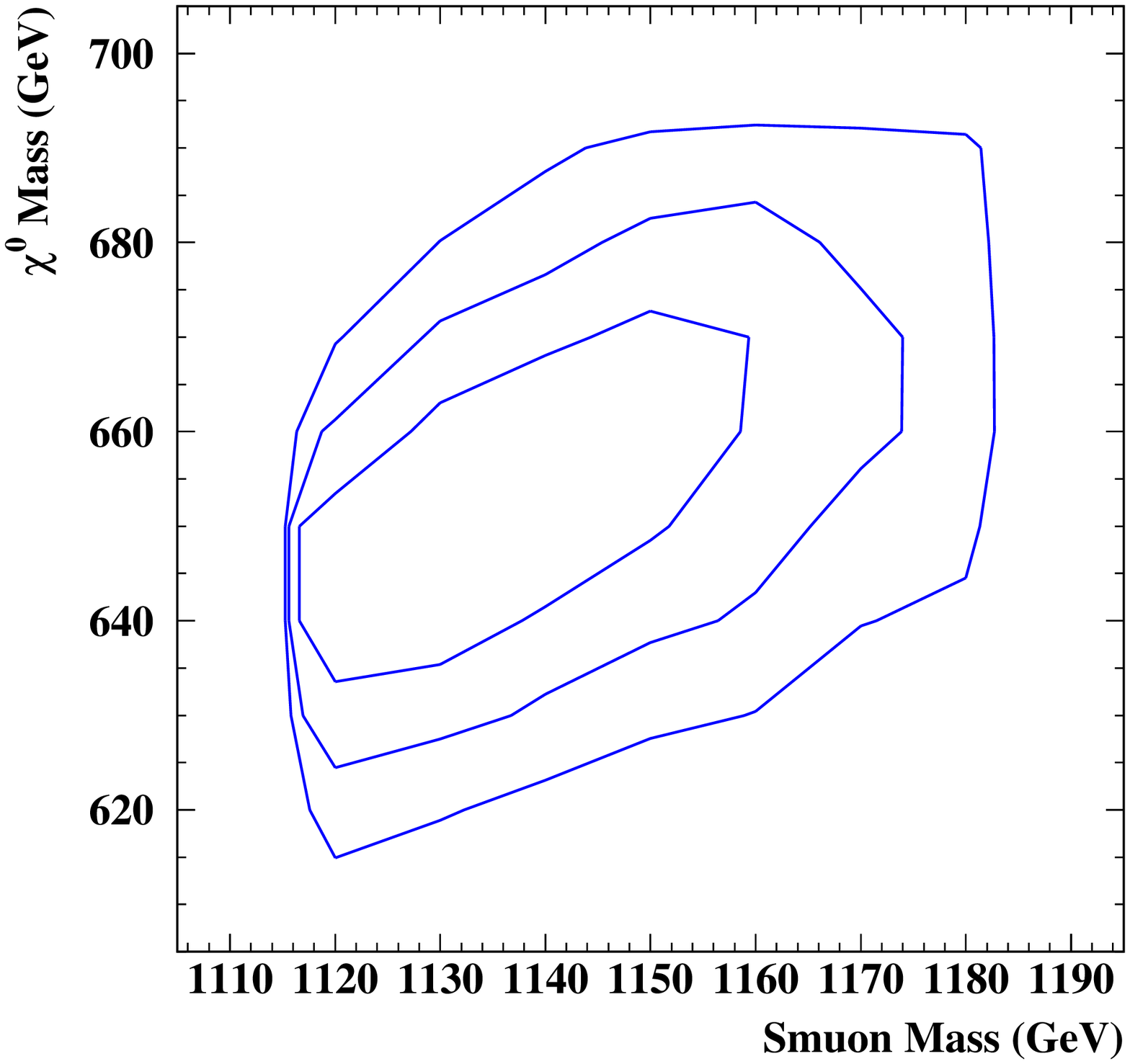,width=2.5in,angle=0}}
\caption{Left: The spectrum of muon momenta in a simulation of
$\tilde \mu \to \chi \mu$ decay at CLIC at $\sqrt{s} = 3$~TeV. Right: Likelihood
contours in the $(m_{\tilde \mu}, m_\chi)$ plane~\protect\cite{CLICphys}.}
\label{fig:sparticles}
\end{figure}

Examples of possible CLIC sparticle measurements are shown in 
Fig.~\ref{fig:sparticles}. In the left panel, we see that CLIC would be able to 
measure well the dilepton spectrum in $\chi^0_2 \to \chi \ell^+ \ell^-$ decay,
locating the endpoint with an accuracy of 2\%. Studies indicate that this
dilepton signature could be measured all along the WMAP strip for
$\tan \beta = 10$, considerably beyond the LHC reach. The right panel
shows how CLIC could measure the smuon decay spectrum, yielding an
accuracy of 2.5\% for the smuon mass and 2\% for the LSP mass in this
particular example~\cite{CLICphys}.

These few examples indicate that, if the LHC discovers supersymmetry,
CLIC could complete the spectrum, and would be able to
make many novel and detailed measurements. By comparing accurate
measurements of the squark and slepton masses, for example, CLIC
may be able to cast light on the mechanism of supersymmetry breaking.
Such measurements may thereby open an interesting window on string physics~\cite{CLICphys}.

The above examples assumed a neutralino LSP, but an alternative is  a
gravitino LSP. In this case, if the scale at which supersymmetry breaking
originates is large, the next-to-lightest sparticle (NLSP) would be metastable, 
since it would decay via gravitational-strength interactions. The NLSP need
not be neutral in such a scenario, and a metastable charged NLSP would
have many interesting signatures.

\begin{figure}[htb]
\centering
{\epsfig{file=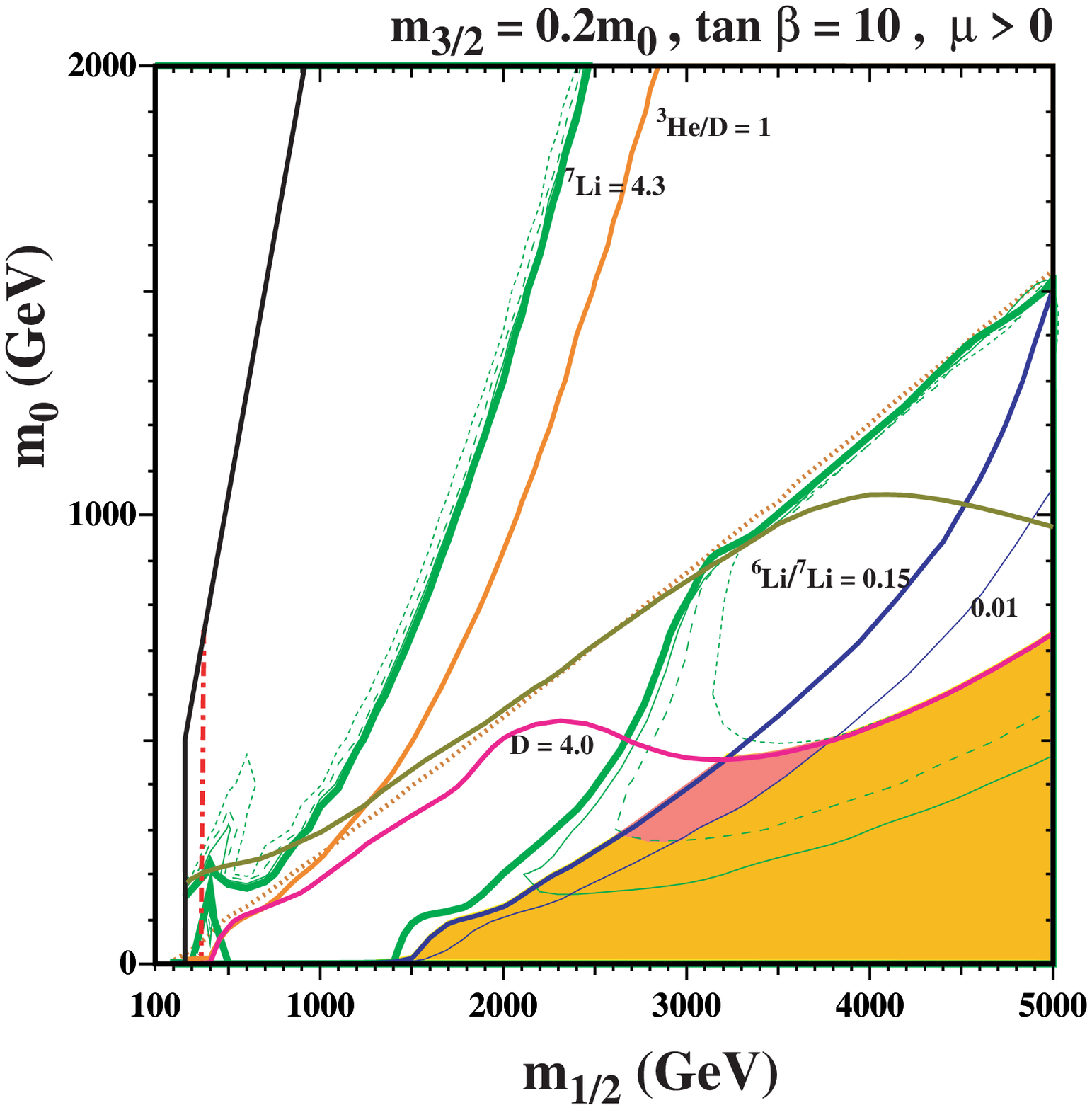,width=2.1in,angle=0}
\epsfig{file=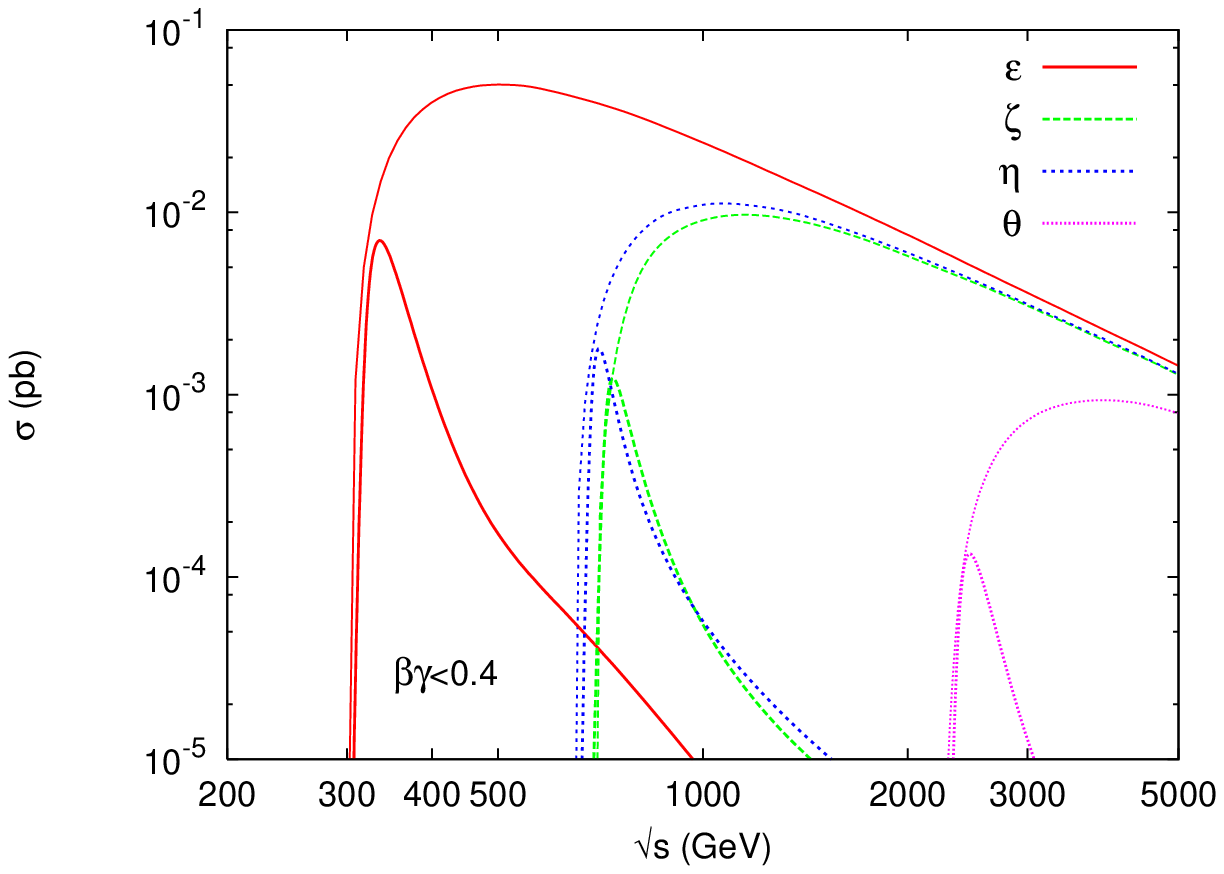,width=2.9in,angle=0}}
\caption{Left: The $(m_{1/2}, m_0)$ plane in a CMSSM scenario with gravitino LSP,
with the BBN constraints shown by solid lines, the resulting BBN-compatible region
being shaded yellow (light) and the region with a preferable $^6$Li/$^7$Li
ratio is shaded pink (darker)~\protect~\cite{gravitinoLSP}. Right: The cross sections for $e^+ e^- \to 
{\bar {\tilde \tau_1}} {\tilde \tau_1}$ production in four benchmark scenarios~\protect\cite{Bench3,Cakir},
showing also the cross sections for producing stoppable staus with $\beta \gamma < 0.4$.
Scenario $\theta$ is from the pink `Lithium sweet spot' in the left panel~\protect\cite{Cakir}.}
\label{fig:gravitino}
\end{figure}

The left panel of Fig.~\ref{fig:gravitino} displays the $(m_{1/2}, m_0)$ plane for one example
of a scenario with a gravitino LSP, with a mass $m_{3/2} = 0.2 m_0$~\cite{gravitinoLSP}. The
NLSP decays could in principle mess up the agreement between Big-Bang 
Nucleosynthesis (BBN) calculations of the light-element abundances and
astrophysical observations. Respecting these constraints, and
incorporating the important effects of stau bound states~\cite{Pospelov}, one is forced
into the light (yellow) shaded region of the left panel of Fig.~\ref{fig:gravitino}. In fact, the
BBN calculations do not agree perfectly with the measured $^6$Li/$^7$Li
ratio, and the darker (pink) shaded region in Fig.~\ref{fig:gravitino} is that
where stau NLSP decays actually {\it improve} the BBN calculations~\cite{gravitinoLSP}.

In order to study the CLIC capabilities for sparticle measurements in
gravitino LSP scenarios with a stau NLSP, we have considered four 
benchmark scenarios, three with relatively light staus detectable at the
LHC~\cite{Bench3}, and one ($\theta$) chosen inside this Lithium `sweet spot'~\cite{Cakir}. 
The total cross sections for $e^+ e^- \to {\bar {\tilde \tau_1}} {\tilde \tau_1}$ production
in these four benchmark scenarios are shown in the right panel of 
Fig.~\ref{fig:gravitino}.

Also shown there are the cross sections for producing slow-moving
staus with $\beta \gamma < 0.4$, which decrease rapidly as $E_{CM}$
increases~\cite{Cakir}. The interest in these events is that such slow staus would stop
in a typical experimental calorimeter. They would then decay later into a
gravitino and a tau, and the decay lifetime and tau energy would provide
valuable information about the mass of the gravitino and the mechanism
of supersymmetry breaking.

\begin{figure}[htb]
\centering
\epsfig{file=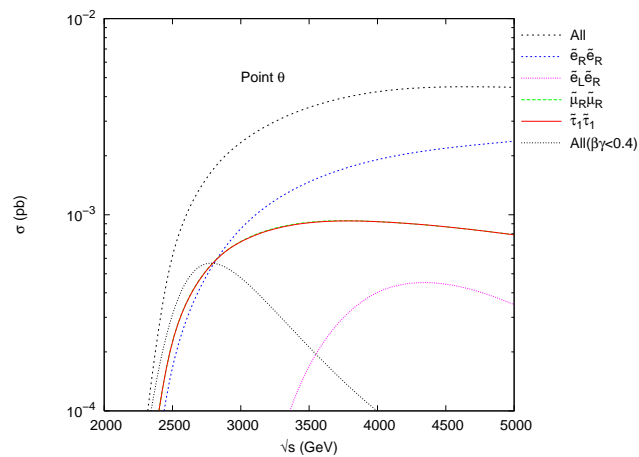,width=3.5in,angle=0}
\caption{Compilation of the principal $e^+ e^-$ annihilation cross sections in
scenario $\theta$. Comparison with Fig.~\protect\ref{fig:gravitino}
shows that the total cross section for stoppable staus
with $\beta \gamma < 0.4$ is considerably larger than that from
$e^+ e^- \to {\bar {\tilde \tau_1}} {\tilde \tau_1}$ alone~\protect\cite{Cakir}.}
\label{fig:all}
\end{figure}

In fact, since all heavier sparticles decay into gravitinos via staus in such a
scenario, the total cross section for stau production is much larger than the
simple pair-production cross section shown in  the right panel of 
Fig.~\ref{fig:gravitino}, as also is the cross section for stoppable stau
production. This is shown in Fig.~\ref{fig:all}, where we see that, e.g., the
total cross section for stoppable stau production at $E_{CM} = 3$~TeV
is about 30 times larger than that from $e^+ e^- \to {\bar {\tilde \tau_1}} {\tilde \tau_1}$
alone.

We conclude that CLIC would be good for detecting and measuring
supersymmetry also in gravitino LSP scenarios, which might even require
relatively heavy spectra.

\section*{8 - Other possible CLIC Physics}

The second-favourite option for new physics beyond the Standard Model
may be extra dimensions. They could rewrite (at least) the hierarchy
problem, might provide a dark matter candidate, and could help in the
unification of the fundamental interactions. Moreover,
extra dimensions are also essential in string theory.

\begin{figure}[htb]
\centering
{\hspace{0.75in}
\epsfig{file=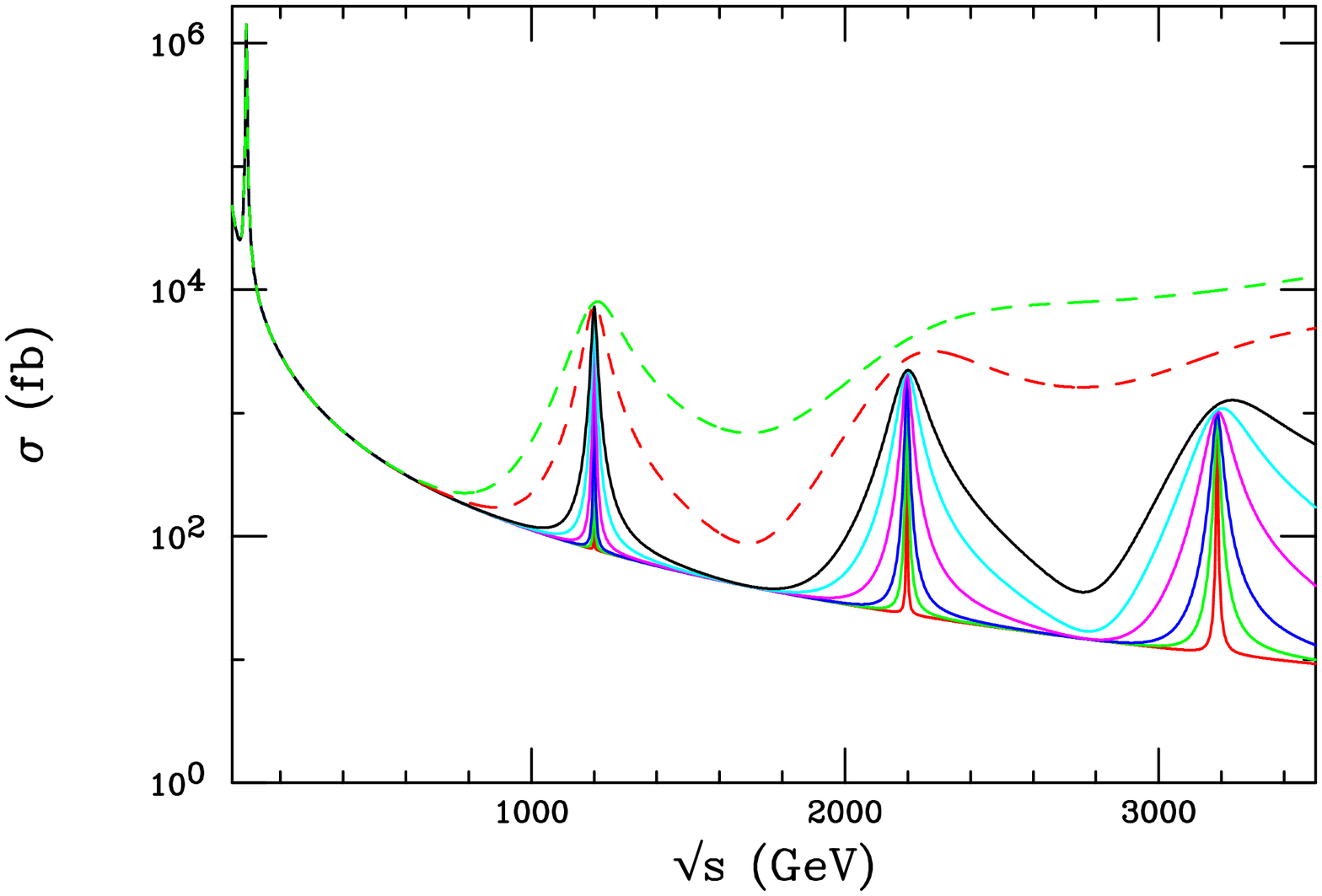,width=4in,angle=90}
\hspace{-0.75in}}
\caption{The spectrum of Kaluza-Klein graviton excitations produced in a 
Randall-Sundrum model in the process $e^+e^-\to \mu^+\mu^-$, showing
different possibilities for their decay widths~\protect\cite{CLICphys}.}
\label{fig:KK}
\end{figure}

They might show up via Kaluza-Klein excitations of Standard Model
particles, which would appear as direct-channel resonances in $e^+ e^-$
annihilation. The high energy offered by CLIC might even provide the
opportunity to observe more than one excitation, as seen in Fig.~\ref{fig:KK}.
It has been shown that CLIC could measure the mass of a $Z^\prime$
boson with an accuracy of 0.01\%, and its width with an accuracy of 0.4\%~\cite{CLICphys}.

In some extra-dimensional scenarios, gravity leaks out from four
dimensions, and may become strong at some energy $\sim 1$~TeV
accessible to the LHC. In this case, CLIC might be able to produce
microscopic black holes. These would decay very quickly into energetic 
quarks, leptons, photons and neutrinos. Such black-hole events would
be easy to distinguish from Standard Model backgrounds~\cite{CLICphys}.

%Table~\ref{tab:comparisons} compares the sensitivities to these and other
%examples of possible new physics of the LHC, ILC, SLHC and CLIC.
%We see that CLIC has particular advantages in the searches for
%massive sleptons, a new $Z^\prime$ gauge boson, extra space dimensions,
%strong $WW$ scattering and measuring the triple-gauge-boson couplings.

\section*{9 - Conclusions}

CLIC will provide unique, high-precision physics at the energy frontier.
In the TeV energy range, it provide beamstrahlung and backgrounds 
similar to those provided by the ILC. Futhermore,
detailed experimental simulations have shown that the beamstrahlung 
and other backgrounds at CLIC would not present insurmountable
obstacles to exploiting fully the higher centre-of-mass energies made available by CLIC.
Several specific examples given above show that CLIC will be able to make
accurate measurements at high energies. The high energy offered by CLIC
will added value to studies of the physics of a light Higgs boson, and
provide unique access to a heavy Higgs boson. CLIC would also have
advantages in studies of supersymmetry or extra dimensions, should they
appear at the LHC. If the new physics beyond the Standard Model has a
relatively low threshold, CLIC will provide unique insight into the heavier
states that may help distinguish between models. On the other hand, if
the new physics is heavy, CLIC may be the only place to study it with
precision.

The future course of high-energy physics will be determined by the LHC,
and we do not know what it will find. However, all the scenarios that have
been studied would best be explored by a high-energy $e^+ e^-$ collider.
Since we do not know the LHC threshold, the world community should 
have available the widest possible technology choice when LHC results 
appear. The ILC technology is more mature than that of CLIC, but the
latter offers more flexibility in energy. Until the time comes to choose, the
CLIC and ILC teams are working together, for example in studies of
positron sources, damping rings, beam dynamics, beam delivery,
interaction regions, detectors and costing. The aim of the CLIC team is
to determine the feasibility of the CLIC technology by the end of 2010,
around the time when the first LHC physics results will become
available, and the time comes for the particle physics community to
decide its next step in collider physics.

\section*{Acknowledgements}

Engin Ar{\i}k was a pillar of Turkish high-energy physics: she and her colleagues
are sorely missed. As discussed in the talks at this meeting, Engin was active in many
different areas of particle physics, and I personally was often grateful for her energy
and advice on relations between Turkey and CERN. I thank her colleagues for
inviting me to this meeting, and wish them good luck and good fortune in building
upon her manifold legacy.

\end{document}